\pdfoutput=1
\documentclass[11pt,twoside,a4paper,cmspaper,final,collab]{cms-tdr}
\def\svnVersion{0e38105}\def\svnDate{2026/09/10}\def\cmsCernNoTag{CERN-EP-2026-080}\def\cmsCernDate{\today}\def\cmsMessage{Published in Physical Review Letters as \href{http://dx.doi.org/10.1103/gwj9-5bxm}{\doi{10.1103/gwj9-5bxm}.}}
\begin{document}\cmsNoteHeader{HIG-24-003}

\ifthenelse{\boolean{cms@external}}{%
    \providecommand{\cmsLeft}{upper\xspace}
    \providecommand{\cmsRight}{lower\xspace}
    \providecommand{\cmsTable}[1]{\resizebox{\columnwidth}{!}{\scriptsize#1}}
     \providecommand{\cmsTableu}[1]{\scriptsize#1}
}{
    \providecommand{\cmsLeft}{upper\xspace}
    \providecommand{\cmsRight}{lower\xspace}
     \providecommand{\cmsTable}[1]{{#1}}
     \providecommand{\cmsTableu}[1]{\resizebox{\columnwidth}{!}{#1}}
}

\newlength\cmsTabSkip\setlength{\cmsTabSkip}{1ex}

\newlength{\cmsFigWidth}
\ifthenelse{\boolean{cms@external}}{\setlength{\cmsFigWidth}{\columnwidth}}{\setlength{\cmsFigWidth}{0.6\textwidth}}

\ifthenelse{\boolean{cms@external}}{%
    \newcommand{\appname}{End Matter}
    \def\appendixname{\appname}
    \providecommand{\EndMatter}{ in the \appname\xspace}
    \providecommand{\suppMaterial}{Supplemental Material~\cite{suppMatBib}}
}{
    \providecommand{\EndMatter}{ in the \nameref{endmatter}\xspace}
    \providecommand{\suppMaterial}{\nameref{suppmat}\nocite{suppMatBib}\xspace}
}

\newcommand{\Pj}{\ensuremath{\mathrm{j}}\xspace}
\newcommand{\klambda}{\ensuremath{\kappa_{\lambda}}\xspace}
\newcommand{\kWW}{\ensuremath{\kappa_{2\PW}}\xspace}
\newcommand{\kZZ}{\ensuremath{\kappa_{2\PZ}}\xspace}
\newcommand{\kVV}{\ensuremath{\kappa_{2\PV}}\xspace}
\newcommand{\WH}{\ensuremath{\PW\PH}\xspace}
\newcommand{\VVH}{\ensuremath{\PV\PV\PH}\xspace}
\newcommand{\VVHH}{\ensuremath{\PV\PV\PH\PH}\xspace}
\newcommand{\WWHH}{\ensuremath{\PW\PW\PH\PH}\xspace}
\newcommand{\ZZHH}{\ensuremath{\PZ\PZ\PH\PH}\xspace}
\newcommand{\Htobb}{\ensuremath{\PH\to\bbbar}\xspace}
\newcommand{\Vtoqq}{\ensuremath{\PV\to\mathrm{qq}}\xspace}
\newcommand{\Xtobb}{\ensuremath{X\to\bbbar}\xspace}
\newcommand{\Xtoqq}{\ensuremath{X\to\mathrm{qq}}\xspace}
\newcommand{\Wjets}{\ensuremath{\PW+\text{jets}}\xspace}
\newcommand{\Zjets}{\ensuremath{\PZ+\text{jets}}\xspace}
\newcommand{\MSD}{\ensuremath{m_{\mathrm{SD}}}\xspace}
\newcommand{\MPN}{\ensuremath{m_{\mathrm{PN}}}\xspace}
\newcommand{\Mjj}{\ensuremath{m_{\Pj{}\Pj}}\xspace}
\newcommand{\detajj}{\ensuremath{\Delta\eta_{\Pj\Pj}}\xspace}
\newcommand{\absdetajj}{\ensuremath{\abs{\detajj}}\xspace}
\newcommand{\std}{\ensuremath{\,\sigma}\xspace}
\newcommand{\DeepJet}{\textsc{DeepJet}\xspace}
\newcommand{\ParticleNet}{\textsc{ParticleNet}\xspace}
\newcommand{\muF}{\ensuremath{\mu_{\mathrm{F}}}\xspace}
\newcommand{\muR}{\ensuremath{\mu_{\mathrm{R}}}\xspace}
\newcommand{\ST}{\ensuremath{\mathrm{S}_{\mathrm{T}}}\xspace}

\cmsNoteHeader{HIG-24-003}

\title{Search for associated production of a Higgs boson and two vector bosons via vector boson scattering at \texorpdfstring{$\sqrt{s} = 13\TeV$}{sqrt(s) = 13 TeV}}

\date{\today}

\abstract{
A search for Higgs boson (\PH) production in association with two vector bosons ($\PV = \PW$, $\PZ$) via vector boson scattering (VBS) is presented using proton-proton collision data collected at $\sqrt{s}=13\TeV$ by the CMS experiment, corresponding to an integrated luminosity of 138\fbinv. Events containing two forward jets consistent with VBS, a large-radius jet from the decay of a boosted \PH to a pair of \PQb quarks, and 0, 1, or 2 charged leptons coming from \PV decays are selected. The process is excluded at 95\%\CL for observed (expected) values of the \VVHH coupling modifier \kVV outside the interval $0.40 < \kVV < 1.60$ ($0.34 < \kVV < 1.66$), assuming standard model values for all other couplings, thus establishing a novel probe of the \VVHH interaction. Constraints are also set on the individual \kWW and \kZZ coupling modifiers, and on the allowed region in the \kWW-\kZZ plane.
}

\hypersetup{%
pdfauthor={CMS Collaboration},%
pdftitle={Search for associated production of a Higgs boson and two vector bosons via vector boson scattering at sqrt(s) = 13 TeV},%
pdfsubject={CMS},%
pdfkeywords={CMS, Higgs, vector boson scattering}}

\maketitle

\begin{figure}[htbp]
\centering
\includegraphics[width=0.49\columnwidth]{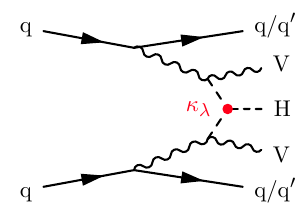}
\includegraphics[width=0.49\columnwidth]{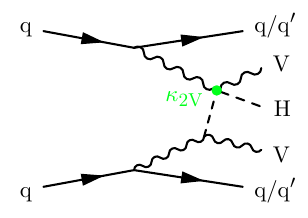}
\caption{
Examples of tree-level Feynman diagrams for the production of \VVH via VBS, with dependencies on the \PH self-coupling (left) and the \VVHH quartic coupling (right). The corresponding vertices are denoted by the \klambda and \kVV modifiers, respectively.
}\label{fig:feynman}
\end{figure}

The discovery of the Higgs boson (\PH) by the ATLAS and CMS Collaborations at the CERN LHC in 2012~\cite{Aad:2012tfa,Chatrchyan:2012ufa,CMS:2013btf} marked a milestone in our understanding of fundamental interactions.
Since then, extensive efforts have focused on measuring its properties and couplings. The \PH self-coupling and the \VVHH ($\PV = \PW$, $\PZ$) quartic coupling, the modifiers of which with respect to the standard model (SM) are parametrized as \klambda and \kVV~\cite{Heinemeyer:2013tqa,deFlorian:2016spz}, respectively, have so far been probed only in searches for \PH pair production~\cite{CMS:comb25,ATLAS:2024ish}.
An alternative path to access both these couplings is offered by \PH production in association with a pair of vector bosons (\VVH) via vector boson scattering (VBS).
This is illustrated in Fig.~\ref{fig:feynman}, where two of the leading order (LO) Feynman diagrams are shown.

The VBS \VVH production cross section at $\sqrt{s}$ = 13\TeV is predicted to be $1.99\unit{fb}$ at LO~\cite{Alwall:2014hca},
considering all the relevant processes.
However, deviations in \kVV from the SM value of 1 can significantly enhance the cross section and boost the final-state objects, thereby increasing the experimental sensitivity to this process. This makes the search for VBS \VVH production synergistic with \PH pair production searches,
which constrain \kVV within the range of $\left[0.64, 1.40\right]$ using CMS data collected at $\sqrt{s}=13\TeV$~\cite{CMS:comb25}. ATLAS data constrain the same parameter in the range $\left[0.6, 1.5\right]$~\cite{ATLAS:2024ish}.
More recently, a CMS search for \PH pair production in final states with four \PQb quarks constrained \kVV in the range $\left[0.74, 1.30\right]$, using improved \PQb tagging algorithms and combining data collected at 13 and 13.6\TeV~\cite{HIG-24-010}.
The VBS \VVH production is also sensitive to the individual \WWHH and \ZZHH couplings, with modifiers \kWW and \kZZ are only loosely constrained by the CMS search for $\PV\PH\PH$ production~\cite{CMS:2024fkb}.
Deviations in \klambda determine less pronounced changes in cross section and kinematics. Consequently, the exploration of the \PH self-coupling is left to future work, and $\klambda=1$ is assumed in this search.

This Letter reports on a search for \VVH production via VBS in proton-proton collision data at $\sqrt{s}=13\TeV$ collected by the CMS experiment at the LHC in 2016--2018, corresponding to an integrated luminosity of $138\fbinv$. The search is performed in several channels: all-hadronic, single-lepton, and dilepton, depending on the number of reconstructed leptons ($\Pell$) from the \PW or the \PZ boson decays.
Dilepton channels target final states containing electrons and muons with charges of the same or opposite sign (SS or OS). Events with one hadronically decaying $\PGt$ lepton are also accepted in the SS channel.
Final states where the \PH decays into a pair of \PQb quarks are studied.
A large Lorentz boost is required for the decay products of the \PH and the vector bosons.
Therefore, the \PH and at least one vector boson decaying hadronically, if present, are reconstructed as single large-radius jets.
All channels utilize the characteristic VBS signature of two additional jets with a large pseudorapidity ($\eta$) separation.

The CMS apparatus~\cite{CMS:2008xjf,CMS:2023gfb} is a multipurpose, nearly hermetic detector, designed to trigger on~\cite{CMS:2020cmk,CMS:2016ngn,CMS:2024aqx} and identify electrons, muons, photons, and hadrons~\cite{CMS:2020uim,CMS:2018rym,CMS:2014pgm}.
The CMS ``particle-flow'' (PF) algorithm~\cite{CMS:2017yfk} combines the information provided by the all-silicon tracker and by the crystal electromagnetic (ECAL) and brass-scintillator hadron calorimeters (HCAL), operating inside a 3.8\unit{T} superconducting solenoid, with data from the gas-ionization muon detectors embedded in the flux-return yoke outside the solenoid, to reconstruct individual particles, which are further used to build \PGt leptons, jets, and missing transverse momentum~\cite{CMS:2018jrd,CMS:2016lmd,CMS:2019ctu}.
The primary vertex (PV) is chosen as the vertex corresponding to the hardest scattering in the event, evaluated using tracking information alone~\cite{CMS-TDR-15-02}.

Muon candidates are reconstructed within $\abs{\eta} < 2.4$ by combining silicon tracker and muon chamber data~\cite{CMS:2018rym}.
Muons must meet quality criteria based on tracker and muon detector hits, track properties, and impact parameters relative to the PV.
Electron candidates within $\abs{\eta} < 2.5$ are reconstructed by matching silicon tracker tracks to ECAL energy clusters~\cite{CMS:2020uim}.
They must meet identification criteria based on shower cluster shape, track-cluster matching, HCAL energy fraction, and track compatibility with the PV.
Electrons in the ECAL barrel-endcap transition region ($1.44 < \abs{\eta} < 1.57$) or originating from photon conversions are discarded.
The lepton isolation is computed by summing the transverse momenta (\pt) of PF candidates originating from the PV within a cone of $\Delta R = \sqrt{\smash[b]{(\Delta\eta)^2+(\Delta\phi)^2}} = 0.4\,(0.3)$ around the muon (electron), where $\phi$ is the azimuthal angle, and is corrected for the contribution of neutral particles
from additional interactions in the same or nearby bunch crossings (pileup)~\cite{CMS:2018rym,CMS:2020uim,CMS-DP-2020-021}.
Selections are based on the isolation sum normalized to the \pt of the lepton.
Leptons are classified as ``tight'' or ``loose'' according to the criteria defined in Ref.~\cite{CMS:2020mpn} in all channels, except for the SS dilepton channel, where criteria similar to those defined in Refs.~\cite{CMS:2018rym,CMS:2020uim,CMS-DP-2020-021} are used.
Hadronically decaying $\PGt$ leptons are identified by the \textsc{DeepTau} algorithm~\cite{CMS:deeptau2022}.

Jets are clustered from PF candidates using the anti-\kt algorithm~\cite{Cacciari:2008gp,Cacciari:2011ma}.
Distance parameters of 0.4 and 0.8 are used to reconstruct so-called ``AK4'' and ``AK8'' (or large-radius) jets, respectively.
Jet momenta are corrected to reduce pileup contributions: the charged hadron subtraction~\cite{CMS:2019kuk} and the pileup-per-particle identification~\cite{CMS:2019kuk,Bertolini:2014bba} algorithms
are used for AK4 and AK8 jets, respectively.
Further corrections are derived from simulation and in-situ measurements in data~\cite{CMS:2016lmd}.
Jets affected by detector noise or reconstruction failures are discarded, and pileup jets are suppressed using a multivariate discriminator~\cite{CMS:2019kuk}.
The missing transverse momentum vector, \ptvecmiss, with magnitude \ptmiss, is defined as the negative \ptvec sum of all PF candidates and corrected for AK4 jet energy scale calibrations~\cite{CMS:2019ctu}.

The mass-decorrelated version of the \Xtobb \ParticleNet tagger~\cite{Qu:2019gqs} is used to identify \Htobb candidates with a mass-independent probability score ranging from 0 to 1.
An analogously mass-decorrelated \Xtoqq \ParticleNet tagger identifies hadronic vector boson decays.
To estimate the AK8 jet mass, the ``soft-drop'' (SD) grooming algorithm~\cite{Larkoski:2014wba,Dasgupta:2013ihk,Butterworth:2008iy} with $\beta = 0$ and $z = 0.1$ is applied, yielding the SD mass, \MSD.
The AK8 mass resolution is further improved using the \ParticleNet-regressed mass, \MPN.
For boosted \PH and vector boson reconstruction, AK8 jets with $\abs{\eta}<2.5$, ungroomed mass $> 50\GeV$, and $\MSD > 40\GeV$ are considered.
After selecting AK8 jets, nonoverlapping AK4 jets are used for \PQb and VBS jet identification.
The AK4 jets originating from \PQb quarks within $\abs{\eta}<2.5$ are identified using the \DeepJet~\cite{CMS:2017wtu,Bols:2020bkb,CMS-DP-2018-058} algorithm
and are used to suppress the background arising from \ttbar production.
Jets are classified as \PQb jets based on the \DeepJet output, with loose, medium, and tight working points corresponding to light-flavor quark or gluon jet misidentification rates of 10, 1, and 0.1\%,
and efficiencies of about 92, 75, and 58\%, respectively.
The AK4 jets with $\pt > 30$\GeV and $\abs{\eta} < 4.7$ are considered as VBS jet candidates.

We model VBS \VVH events and calculate the corresponding production cross sections at LO using the \MGvATNLO~v2.6.5~\cite{Alwall:2014hca} Monte Carlo event generator.
The generator is interfaced with \PYTHIA~v8.240~\cite{Sjostrand:2014zea} for the simulation of boson decays, using the dipole recoil scheme to correctly emulate the color structure~\cite{Cabouat:2017rzi}.
To simulate events with different \kVV values, the event reweighting feature of \MGvATNLO~\cite{mattelaer16_261} is employed.

Background simulation is used in the optimization of the analysis, while data-driven methods are used for the purpose of modelling the data in all channels.
Simulated events are also used in the SS dilepton channel to predict the SM background, in conjunction with control samples in data.
The largest source of background in the all-hadronic channel is the production of multijet events through quantum chromodynamics (QCD) processes.
The QCD background is simulated at LO with \MGvATNLO. Other major background processes are \ttbar and \Wjets production.
The $\ttbar$ background~\cite{Frixione:2007nw} is simulated at next-to-LO (NLO) precision in perturbative QCD with the \POWHEG~v2.0 program~\cite{Nason:2004rx,Frixione:2007vw,Alioli:2010xd}.
The \Wjets background is simulated with \MGvATNLO at LO, with up to four additional partons, merged at the parton shower level via the MLM matching scheme~\cite{Alwall:2007fs}.
Other sources of background include diboson production, $s$-, $t$-channel and $\PQt\PW$ single top quark production, electroweak production of a \PW boson, quark-initiated \WH, \ttbar~$+$~vector bosons, and \Zjets production.

The generated matrix-element-level events are interfaced with \PYTHIA~v8.240 to model parton showering and hadronization, as well as the underlying event.
The generation of all samples is based on the CP5 tune~\cite{tuneCP5Paper} for \PYTHIA and the NNPDF~v3.1 next-to-NLO parton distribution functions (PDFs)~\cite{Ball:2014uwa,Ball:2017nwa}.
Simulated pileup interactions are added to the hard-scattering process, with a multiplicity matched to the pileup profiles of the data. The detector response is modeled with \GEANTfour~\cite{GEANT4:2002zbu}.

The search is performed across six channels, defined according to the lepton and jet multiplicities in the event.
Events in the all-hadronic channel are required to contain exactly zero loose leptons, while in the single-lepton (dilepton) channel they must contain one (two) tight leptons.
The all-hadronic channel is further split into two categories, semi-boosted and fully boosted, which require exactly two and more than two AK8 jets, respectively.
The dilepton events are split into the OS and SS channels. Events in the OS dilepton channels are then categorized according to the dilepton mass ($m_{\Pell\Pell}$), requiring that it is either consistent with that of a \PZ boson (OS \PZ), or not (OS $\PW\PW$).
All channels are orthogonal by construction except for the SS dilepton channel. For this channel the orthogonality is explicitly verified after the event selections are applied, because of the different lepton identification criteria employed.

In the all-hadronic channels, events of interest are collected using triggers based on the scalar \pt sum of all jets ($H_\mathrm{T}$), with thresholds of 800 or 900\GeV for 2016, and 1050\GeV for 2017 and 2018.
Events are split into the fully boosted and semi-boosted channels based on the number of AK8 jets with $\pt > 300\GeV$.
In the fully boosted channel, the \Htobb and both the vector boson candidates are reconstructed as AK8 jets.
In the semi-boosted channel, one vector boson candidate is reconstructed as a pair of AK4 jets.
In both channels, a pair of AK4 VBS jet candidates is also required: in the case of multiple pairs, the one with the largest absolute pseudorapidity difference (\absdetajj) is selected.

In the single-lepton channel, single-lepton triggers require isolated leptons with $\pt > 24$~(27)\GeV for muons in 2016 and 2018 (2017), and $\pt > 27$~(32)\GeV for electrons in 2016 (2017--2018).
Events must contain exactly one tight lepton with $\pt > 40\GeV$ and no additional loose leptons.
Two AK8 jets with $\pt > 250\GeV$ are required and identified as the \Htobb and the hadronically decaying vector boson candidates.
The VBS jets are chosen as the highest-momentum AK4 jets in opposite $\eta$ regions.
If this criterion fails, the two highest-momentum jets are selected.

In the dilepton channels, single-lepton triggers are used for OS $\PW\PW$ events. These are used in combination with dilepton triggers for OS \PZ events, with \pt thresholds of 17 and 8\GeV for muons, and 23 and 12\GeV for electrons.
The SS dilepton channel also uses mixed muon-electron triggers, requiring a muon (electron) with $\pt > 23$~(12)\GeV and an electron (muon) with $\pt > 23$~(8)\GeV.
Exactly two tight leptons are required, with no additional loose leptons.
In the OS \PZ channel, the two leptons must have an invariant mass $m_{\Pell\Pell}$ consistent with that of a \PZ boson, requiring $86 < m_{\Pell\Pell} < 96\GeV$, while the same window is vetoed in the OS $\PW\PW$ channel.
The \Htobb candidate is reconstructed as an AK8 jet in all channels.
In the OS \PZ channel, one vector boson decays hadronically and is also reconstructed as an AK8 jet.
The VBS jet pair is chosen as the pair of AK4 jets with the largest $\absdetajj$ in both the OS channels, while the same selection as in the single-lepton channel is used in the SS channel.

An AK4 \PQb jet veto is applied in the single-lepton and dilepton channels, to reduce the \ttbar background contribution.
Additional selection criteria on event properties are also applied, as summarized in Tables~\ref{tab:fhfb_presel}--\ref{tab:ss_presel}\EndMatter.

A multivariate analysis is subsequently performed in all channels.
Backgrounds are estimated from control samples in data using a matrix (``ABCD'') method~\cite{CDF:1990kbl}, except for the SS dilepton channel.
The background in the signal region (A) is predicted from data in control regions B, C, and D, defined by inverting the criteria used to select the VBS jets (B), the \VVH system (C), or both (D).
This choice of the ABCD axes is motivated by the negligible correlation expected in background events between the properties of the \VVH system and the VBS jets.

The all-hadronic and single-lepton channels use the method suggested in Ref.~\cite{Kasieczka:2020pil} to train a deep neural network (DNN) orthogonal to the discriminating features used to select the VBS jet pair.
The output of the DNN and the VBS selection features are then used as the two ABCD axes.
Different DNNs are trained in each channel, using simulated events to separate signal in the $\kVV=2$ benchmark scenario from all background sources.
This benchmark is used as it is expected to be close to the final exclusion limit.
In the all-hadronic fully boosted topology, the DNN uses as input the \pt, $\eta$, $\phi$, and \MPN of the three selected AK8 jets, along with the invariant mass of the VBS jet pair, \Mjj.
A decorrelation term in the loss function ensures that the DNN output is independent of $\detajj$, and these same quantities are used to define the ABCD axes.
After training, the minimum thresholds used to select on the \Xtobb and \Xtoqq \ParticleNet tagger scores are tightened for both the \PH and the vector boson candidates.
The signal region A is defined by $\absdetajj > 5$ and a DNN output $>0.97$.
In the semi-boosted topology, the DNN is trained using the \pt, $\eta$, $\phi$, and \MSD of the two selected AK8 jets, the four-vectors and the invariant mass \Mjj of the AK4 jet pair forming the second vector boson candidate, and the \pt sum of all objects forming the \VVH system, \ST.
The DNN output is designed to be independent of $\detajj$.
The signal region A is defined by requiring $\absdetajj > 6$, $\Mjj > 1300\GeV$, and a DNN output $> 0.8$.
In the single-lepton channel, the DNN inputs are the \pt and \MPN of the \PH and vector boson AK8 jets, the lepton \pt, the $m_{\Pell\PQb}$ variable, defined as the minimum invariant mass of the lepton and any \PQb jet passing the loose \DeepJet tagger working point, and \ptmiss.
Its output is constructed to be orthogonal to that of a boosted decision tree (BDT) used to select the VBS jet pair, trained on the \pt and $\eta$ of the VBS jet candidates, \detajj and \Mjj.
The signal region A is defined by requiring the DNN and BDT outputs to be $>0.9$ and 0.82, respectively, with the \Xtobb (\Xtoqq) tagger score $> 0.5$ (0.7).

In each dilepton channel, a BDT is trained to separate signal (in the $\kVV=2$ benchmark scenario) from simulated background.
The BDTs are trained using features of the \VVH system. The absence of correlation with the properties of the VBS jets is verified after the training.
In the OS $\PW\PW$ channel, inputs include the \PH candidate \pt, $\eta$, \MSD; the \pt, $\eta$, flavor of each lepton; the $\Delta\eta$, $\Delta R$, invariant mass of the two leptons; the event \ST and \ptmiss.
The signal region A requires $\Mjj > 760\GeV$, $\absdetajj > 4.9$, a BDT score $>0.85$, and an \Xtobb tagger score $>0.7$.
In the OS \PZ channel, the BDT inputs are the \pt and $\eta$ of the \PH and \PZ candidates, and of each lepton; the lepton impact parameter significances and isolation sums; and $m_{\Pell\Pell}$, the invariant mass of the \VVH system, and \ST.
The signal region A is defined by $\Mjj > 750\GeV$, $\absdetajj > 4$, and a BDT score $>0.85$.

The selections used in each channel to define the signal region, or region A, are summarized in Table~\ref{tab:SRs}\EndMatter, where the cuts inverted to define the B, C, D regions are also specified.

Systematic uncertainties in the background predictions are derived from closure tests in data.
An analogous ABCD method is applied in data sidebands, and the prediction is compared to the observed data in the corresponding region A of the sideband.
The uncertainty is estimated conservatively, by taking the relative difference in the worst closure sideband.
Table~\ref{tab:yields}\EndMatter shows the data yields in the regions A, B, C and D, in each channel, alongside the predicted background with its uncertainties, and the benchmark signal expected in region A, for both the $\kVV=2$ and SM scenarios.

In the SS dilepton channel, a BDT is also trained to discriminate signal from background.
The BDT inputs include the \pt and $\eta$ of the \Htobb candidate; the \pt of the leptons; the momentum of the VBS jets and the \pt of the VBS dijet system; and $m_{\Pell\Pell}$, \ptmiss, \ST, and the scalar sum of the \pt of the leptons and \ptmiss.
The signal region is then divided into three bins according to the BDT output score ($\left[0, 0.1\right]$, $\left[0.1, 0.25\right]$, $\left[0.25, 1\right]$).

Results are obtained via binned maximum likelihood fits of signal and background predictions to data~\cite{ParticleDataGroup:2024cfk}, using the CMS \textsc{Combine} tool~\cite{CMS:2024onh}, based on the \textsc{RooFit}~\cite{Verkerke:2003ir} and \textsc{RooStats}~\cite{Moneta:2010pm} frameworks.
In each channel, all the regions are fitted simultaneously, extracting the signal region predictions concurrently.
Systematic uncertainties are included as nuisance parameters of the likelihood function and affect the signal and background normalization, except for the SS dilepton channel where shape variations are used.
Experimental uncertainties in the signal acceptance are considered in the lepton trigger and identification efficiencies, jet energy scale and resolution~\cite{CMS:2016lmd}, \DeepJet \Pb tagging efficiency, \ParticleNet tagger efficiency, and integrated luminosity~\cite{CMS-LUM-17-003,CMS-PAS-LUM-17-004,CMS-PAS-LUM-18-002}.
Theory uncertainties in the signal cross section include renormalization and factorization scales (\muR and \muF, respectively), PDFs, and initial- and final-state radiation modeling. The overall dominant systematic uncertainty, of approximately 25\%, arises from \muF, followed by the uncertainty arising from PDFs, of approximately 10\%.
For the background prediction the statistical and closure systematic uncertainties are considered.
The sensitivity is limited mostly by the statistical uncertainty, given the low data yields in the signal regions.

\begin{figure}[htb]
\centering
\includegraphics[width=\cmsFigWidth]{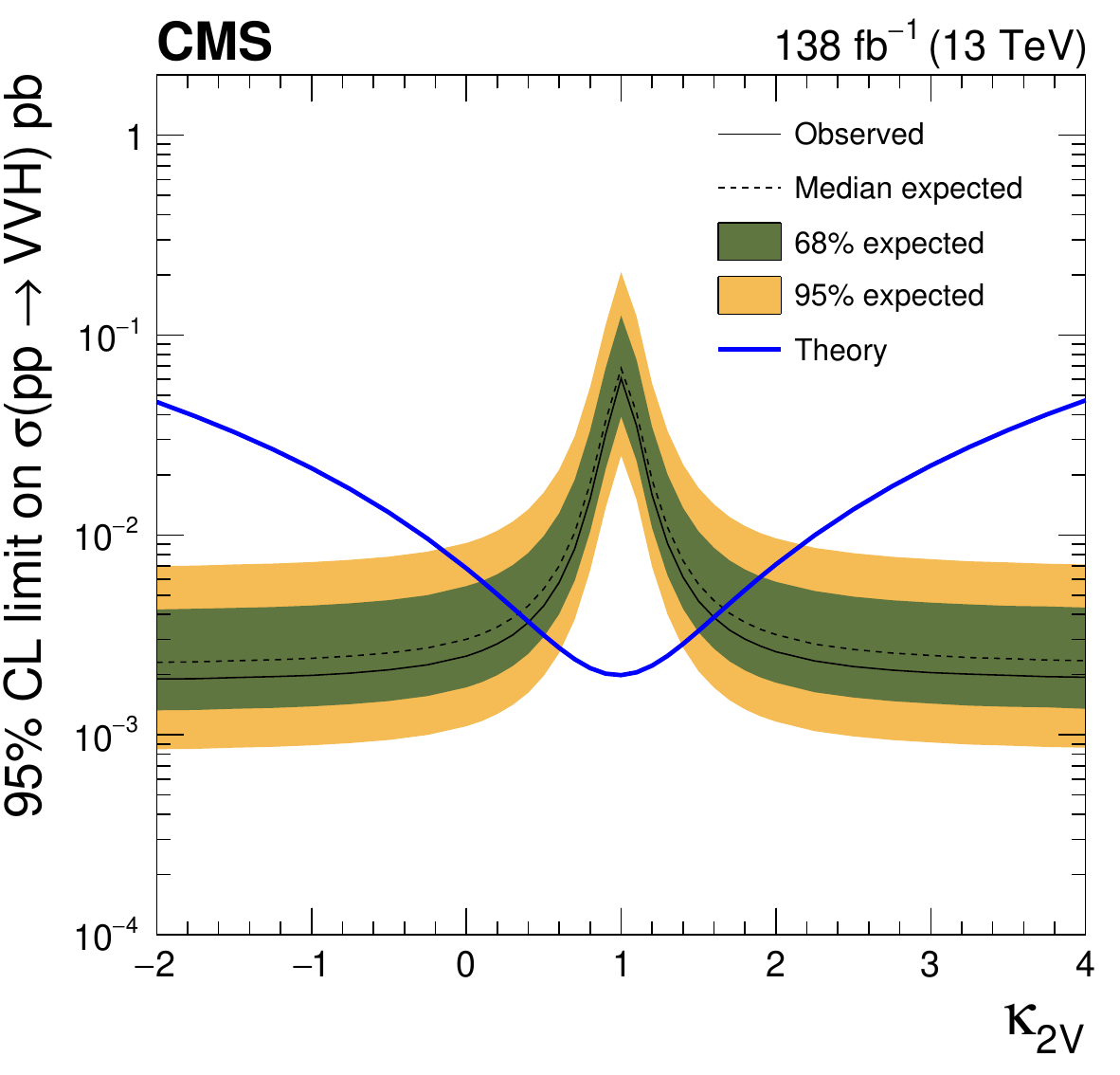}
\caption{
Observed (solid) and expected (dashed) 95\%~\CL constraints on the VBS \VVH production cross section as a function of \kVV, with other couplings fixed to their SM values. The intersections with the predicted cross section (blue) indicate the excluded \kVV ranges.
}
\label{fig:1dscan}
\end{figure}

All channels are combined to obtain upper limits at 95\% confidence level (\CL) on the VBS \VVH production cross section as a function of the \kVV coupling modifier.
Assuming all other \PH couplings are equal to their values in the SM, the VBS \VVH production is excluded at 95\%~\CL for \kVV outside the observed (expected) range of $0.40 < \kVV < 1.60$ ($0.34 < \kVV < 1.66$). The results are shown in Fig.~\ref{fig:1dscan}.

\begin{figure}[htbp]
\centering
\includegraphics[width=\cmsFigWidth]{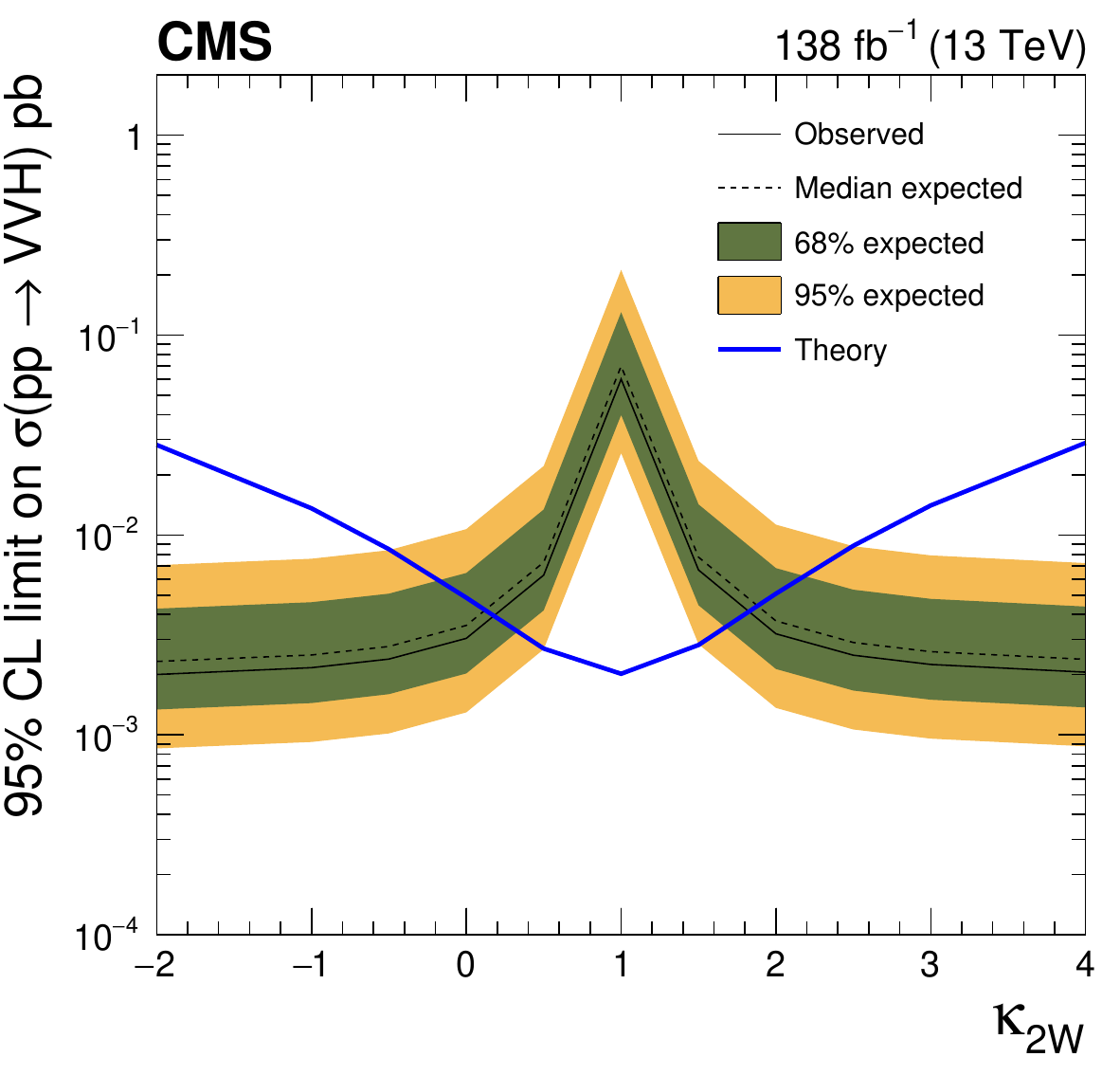}
\includegraphics[width=\cmsFigWidth]{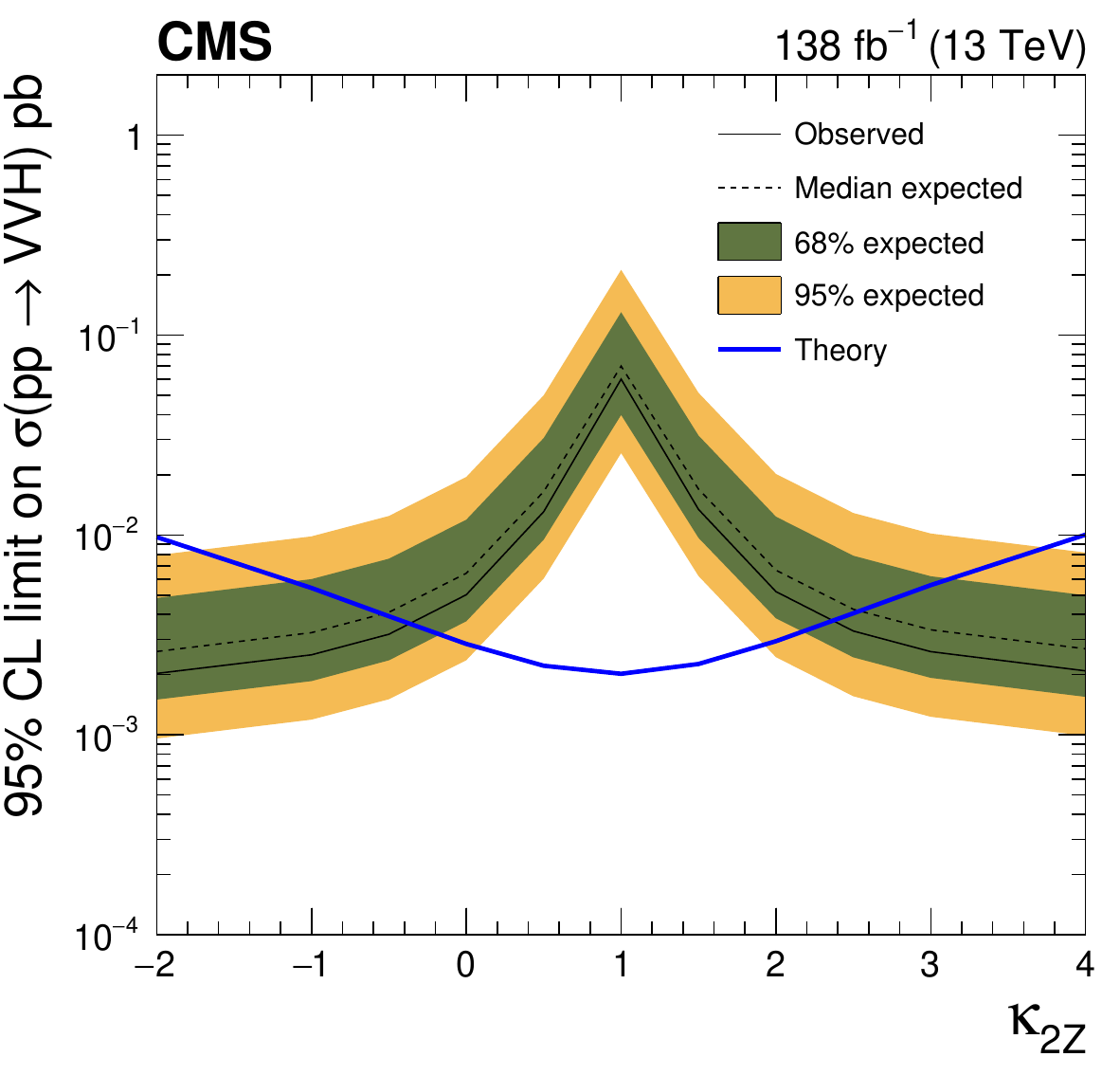}
\caption{
Observed (solid) and expected (dashed) 95\%~\CL constraints on the VBS \VVH production cross section as a function of \kWW (\cmsLeft) and \kZZ (\cmsRight).
The intersections with the predicted cross section (blue) indicate excluded coupling ranges.
}
\label{fig:ilimwz}
\end{figure}

Similarly, the channels are combined to constrain the \kWW and \kZZ parameters, each varied while keeping the other at its SM value, as reported in Fig.~\ref{fig:ilimwz}.
The \kWW and \kZZ coupling modifiers are constrained in the (expected) ranges of $0.17 < \kWW < 1.84$ ($0.11 < \kWW < 1.89$) and $-0.37 < \kZZ < 2.38$ ($-0.54 < \kZZ < 2.54$), respectively.

\begin{figure}[htbp]
\centering
\includegraphics[width=\cmsFigWidth]{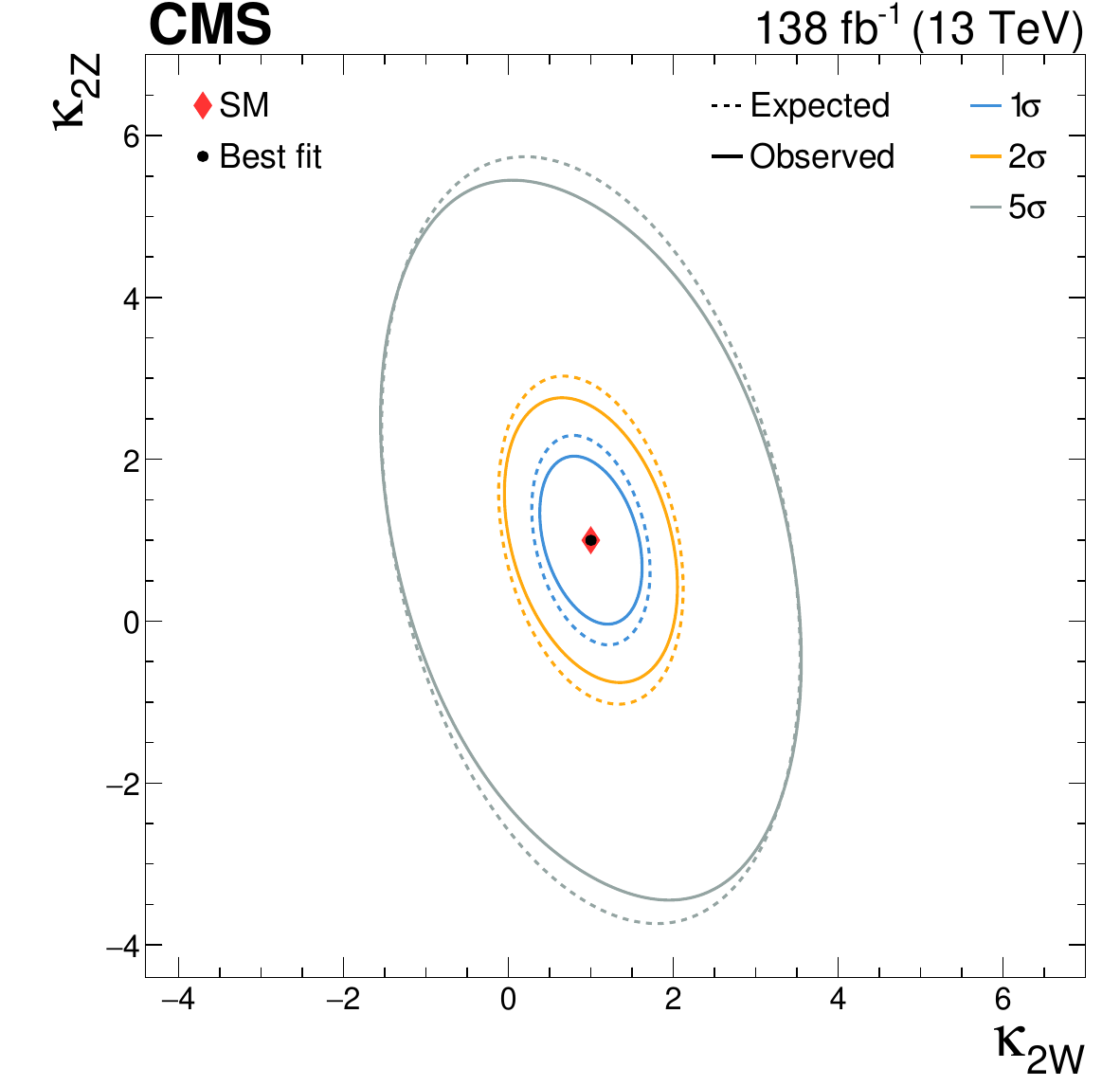}
\caption{
Observed (solid) and expected (dashed) exclusion regions corresponding to 1, 2, and 5 standard deviations (\std), as obtained from a likelihood scan in the two-dimensional \kWW-\kZZ plane.
}
\label{fig:2dscan}
\end{figure}

A two-dimensional likelihood scan in the \kWW-\kZZ plane is also performed to derive constraints in that plane.
The observed (solid line) and expected (dashed line) exclusion regions corresponding to 1, 2 and 5 standard deviations (\std) are shown in Fig.~\ref{fig:2dscan}.

Tabulated results are provided in the HEPData record for this analysis~\cite{hepdata}.
Additional details on the analysis strategies and on the sensitivity achieved in individual channels are provided in the \ifthenelse{\boolean{cms@external}}{\suppMaterial}{\suppMaterial}.

In summary, the first study of \VVH production through vector boson scattering is presented, using proton-proton collision data recorded by the CMS experiment at the LHC in 2016--2018, at $\sqrt{s} = 13\TeV$, corresponding to an integrated luminosity of 138\fbinv.
Selected events are consistent with the presence of two jets originating from VBS and an \PH decaying into a pair of \PQb quarks, reconstructed as a single large-radius jet.
Final states with zero, one, or two charged leptons arising from the decay of two vector bosons are studied.
The VBS \VVH production is excluded at 95\%~\CL for values of the \kVV coupling modifier outside the observed (expected) range of $0.40 < \kVV < 1.60$ ($0.34 < \kVV < 1.66$), assuming all other \PH couplings are equal to their values in the standard model.
The results represent one of the best constraints to date on \kVV using CMS data, complementing the results obtained from searches for \PH pair production, and laying the foundation for future, similar studies.
The \kWW and \kZZ coupling modifiers are also constrained independently in the observed (expected) ranges of $0.17 < \kWW < 1.84$ ($0.11 < \kWW < 1.89$) and $-0.37 < \kZZ < 2.38$ ($-0.54 < \kZZ < 2.54$), respectively.
A two-dimensional scan, determining exclusion regions in the \kWW-\kZZ plane, is also provided, which largely improves on the constraints coming from the CMS search for $\PV\PH\PH$ production.

\begin{acknowledgments}
We congratulate our colleagues in the CERN accelerator departments for the excellent performance of the LHC and thank the technical and administrative staffs at CERN and at other CMS institutes for their contributions to the success of the CMS effort. In addition, we gratefully acknowledge the computing centers and personnel of the Worldwide LHC Computing Grid and other centers for delivering so effectively the computing infrastructure essential to our analyses. Finally, we acknowledge the enduring support for the construction and operation of the LHC, the CMS detector, and the supporting computing infrastructure provided by the following funding agencies: SC (Armenia), BMBWF and FWF (Austria); FNRS and FWO (Belgium); CNPq, CAPES, FAPERJ, FAPERGS, and FAPESP (Brazil); MES and BNSF (Bulgaria); CERN; CAS, MoST, and NSFC (China); MINCIENCIAS (Colombia); MSES and CSF (Croatia); RIF (Cyprus); SENESCYT (Ecuador); ERC PRG and PSG, TARISTU24-TK10 and MoER TK202 (Estonia); Academy of Finland, MEC, and HIP (Finland); CEA and CNRS/IN2P3 (France); SRNSF (Georgia); BMFTR, DFG, and HGF (Germany); GSRI (Greece); MATE and NKFIH (Hungary); DAE and DST (India); IPM (Iran); SFI (Ireland); INFN (Italy); MSIT and NRF (Republic of Korea); MES (Latvia); LMTLT (Lithuania); MOE and UM (Malaysia); BUAP, CINVESTAV, CONACYT, LNS, SEP, and UASLP-FAI (Mexico); MOS (Montenegro); MBIE (New Zealand); PAEC (Pakistan); MSHE, NSC, and NAWA (Poland); FCT (Portugal); MESTD (Serbia); MICIU/AEI and PCTI (Spain); MOSTR (Sri Lanka); Swiss Funding Agencies (Switzerland); MST (Taipei); MHESI (Thailand); TUBITAK and TENMAK (T\"{u}rkiye); NASU (Ukraine); STFC (United Kingdom); DOE and NSF (USA).
\end{acknowledgments}
\bibliography{auto_generated}

\appendix
\ifthenelse{\boolean{cms@external}}{\section{}}{\section{End Matter}\label{endmatter}}
\subsection{Event selection}
\label{endmatter:selection}

This section details the event selections used in each channel before multivariate analyses are performed.
In the all-hadronic channels, events of interest are collected using triggers based on $H_\mathrm{T}$, with thresholds of 800 or 900\GeV for 2016, and 1050\GeV for 2017 and 2018.
Events must contain at least two AK8 jets with $\pt > 300\GeV$, and $\pt > 550$\GeV for the leading one to ensure full trigger efficiency.
Two categories are defined to distinguish fully and semi-boosted events.
In the fully boosted channel, the \Htobb candidate is selected as the AK8 jet with the highest \Xtobb \ParticleNet score ($>0.5$), and the vector boson candidates are selected as the two highest \pt remaining AK8 jets with \Xtoqq scores $>0.3$.
In the semi-boosted channel, exactly two AK8 jets are required.
The \Htobb candidate is selected as the AK8 jet with the highest \Xtobb score ($>0.8$).
One of the two vector bosons is reconstructed as the other AK8 jet in the event, requiring an \Xtoqq minimum score of 0.637, 0.579, and 0.59 for 2016, 2017, and 2018 data taking, respectively.
The second vector boson is reconstructed from two AK4 jets ($\pt > 20$\GeV, $\abs{\eta}<2.5$) chosen as the closest pair in $\Delta R$. Two AK4 VBS jet candidates are also required.
In the case of multiple pairs, the one with the largest \absdetajj is selected as the VBS jet pair.
The selection criteria for the all-hadronic fully boosted and semi-boosted channels are summarized in Tables~\ref{tab:fhfb_presel} and~\ref{tab:fhsb_presel}, respectively.

\begin{table}[!ht]
\centering
\topcaption{Selection criteria used in the all-hadronic fully boosted channel. If selection criteria differ across years, they are quoted as ``2016/2017/2018''.}
\label{tab:fhfb_presel}
\cmsTable{
\begin{scotch}{lc}
Selection             & Description \\
\hline
Trigger               &  $H_\mathrm{T}> 800$ or 900/1050/1050\GeV \\
Loose lepton count    & $=$0  \\
Tight lepton count    & $=$0 \\
AK8 jet threshold     & $\pt>300\GeV$ \\
AK8 jet count         & $>$2 \\
Leading AK8 jet       & $\pt>550\GeV$ \\
\Htobb candidate      & max \Xtobb score AK8 jet \\
                      & \Xtobb score $>0.5$ \\
\Vtoqq candidates     & \pt-leading AK8 jets \\
                      & \Xtoqq score  $> 0.3$ \\
AK4 jet count for VBS & $\geq$2 \\
VBS jet candidates    & max $\absdetajj$ \\
\end{scotch}}
\end{table}

\begin{table}[!ht]
\centering
\topcaption{Selection criteria used in the all-hadronic semi-boosted channel. If the selection criteria differ across years, they are quoted as ``2016/2017/2018''.}
\label{tab:fhsb_presel}
\cmsTable{
\begin{scotch}{lc}
Selection                   & Description \\
\hline
Trigger                     &  $H_\mathrm{T}> 800$ or 900/1050/1050\GeV \\
Loose lepton count          & $=$0  \\
Tight lepton count          & $=$0 \\
AK8 jet threshold           & $\pt>300\GeV$ \\
AK8 jet count               & $=$2 \\
Leading AK8 jet             & $\pt>550\GeV$ \\
\Htobb candidate            & max \Xtobb score AK8 jet \\
                            & \Xtobb score $> 0.8$ \\
Boosted \Vtoqq candidate    & other AK8 jet \\
                            & \Xtoqq score  $> 0.637/0.579/0.59$ \\
AK4 jet for \Vtoqq criteria & $\pt > 20\GeV$, $\abs{\eta}<2.5$ \\
AK4 jet for \Vtoqq count    & $\geq$2 \\
Resolved \Vtoqq candidate   & min $\Delta R$ \\
AK4 jet count for VBS       & $\geq$2 \\
VBS jet candidates          & max $\absdetajj$ \\
\end{scotch}}
\end{table}

In the single-lepton channel, single-lepton triggers require isolated leptons with $\pt > 24$~(27)\GeV for muons in 2016 and 2018 (2017), and $\pt > 27$~(32)\GeV for electrons in 2016 (2017--2018).
Events must contain exactly one tight lepton with $\pt > 40\GeV$ and no additional loose leptons.
Two AK8 jets with $\pt > 250\GeV$ are required. The \Htobb candidate is reconstructed as the AK8 jet with the highest \Xtobb \ParticleNet score, while the hadronic vector boson candidate is the remaining AK8 jet with the highest \Xtoqq score.
The VBS jets are chosen as the highest-momentum jets in opposite $\eta$ regions; if this criterion fails, the two highest-momentum jets are selected. Events with AK4 jets passing the tight \DeepJet working point are rejected.
Jets passing the loose working point are used to define $m_{\Pell\PQb}$, the minimum invariant mass of the lepton and any \PQb jet, used in the multivariate analysis to suppress the \ttbar background.
Furthermore, a selection is applied on $\ST > 1000\GeV$.
The selection criteria for the single-lepton channel are summarized in Table~\ref{tab:sl_presel}.

\begin{table}[!ht]
\centering
\topcaption{Selection criteria used in the single-lepton channel. If the selection criteria differ across years, they are quoted as ``2016/2017/2018''.}
\label{tab:sl_presel}
\cmsTable{
\begin{scotch}{lc}
Selection                         & Description \\
\hline
Trigger                           & Single-muon $\pt > 24/27/24 \GeV$ \\
                                  & Single-electron $\pt > 27/32/32 \GeV$ \\
Sign and flavor of leptons        & \Pepm, \PGmpm \\
Loose lepton count                & $=$1  \\
Tight lepton count                & $=$1 \\
Tight lepton $\pt$                & $\pt>40 \GeV$ \\
AK8 jet threshold                 & $\pt>250 \GeV$ \\
AK8 jet count                     & $\geq$2 \\
\Htobb candidate                  & max $\Xtobb$ score AK8 jet \\
\Vtoqq candidate                  & max $\Xtoqq$ score AK8 jet \\
AK4 jet count for VBS             & $\geq$2 \\
VBS jet candidates                & max momentum, opposite $\eta$ sign \\
AK4 jet for \PQb tagging criteria & $\pt > 20\GeV$, $\abs{\eta}<2.5$ \\
Tight \PQb tagged jet count       & $=$0 \\
Event kinematics                  & $\ST > 1000\GeV$ \\
\end{scotch}}
\end{table}

In the dilepton channels, single-lepton triggers are used for OS $\PW\PW$ events.
These are used in combination with dilepton triggers for OS \PZ events, with \pt thresholds of 17 and 8\GeV for muons, and 23 and 12\GeV for electrons.
In OS $\PW\PW$ events, electrons must have $\pt > 29\,(34)$\GeV in 2016 (2017--2018), and muons $\pt > 26\,(29)$\GeV in 2016 and 2018 (2017).
The two leptons must be oppositely charged and separated with $\Delta R > 0.5$. A \PZ boson mass veto excludes events with $86 < m_{\Pell\Pell} < 96\GeV$. The \PH candidate is selected as the AK8 jet with the highest \Xtobb \ParticleNet score,
$\pt > 250\GeV$, and $70 < \MPN < 160\GeV$.
Events with AK4 jets passing the \DeepJet medium working point are rejected, and the two jets with the largest $\absdetajj$ are selected as VBS jet candidates.
Events must have $\ptmiss > 100\GeV$ and $\ST > 500\GeV$.
In OS \PZ events, the leading muons (electrons) must have $\pt > 19\,(25)$\GeV, and the subleading $\pt > 12\,(18)$\GeV.
The leptons must be oppositely charged, with $86 < m_{\Pell\Pell} < 96\GeV$.
The \Htobb candidate is selected as the AK8 jet with $\pt > 250\GeV$ and highest \Xtobb score ($>$0.2), while the hadronically decaying vector boson candidate is the AK8 jet with $\pt > 200\GeV$ and highest \Xtoqq score ($>$0.6).
The invariant mass $M_{\VVH}$ of the \Htobb candidate, the hadronically decaying vector boson candidate and the lepton pair, and $\ST$ must both exceed 500\GeV.
The VBS jet pair is chosen based on the largest $\absdetajj$, with the minimum requirements of $\Mjj$ $> 200\GeV$ and $\absdetajj > 2$.
Events with AK4 jets passing the \DeepJet medium working point are rejected.
The selection criteria used for OS dilepton events are summarized in Tables~\ref{tab:osww_presel} and~\ref{tab:osz_presel} for the OS $\PW\PW$ and OS $\PZ$ channels, respectively.

\begin{table}[!ht]
\centering
\topcaption{Selection criteria in used the OS $\PW\PW$  dilepton channel. If the selection criteria differ across years, they are quoted as ``2016/2017/2018''.}
\label{tab:osww_presel}
\cmsTable{
\begin{scotch}{lc}
Selection                         & Description \\
\hline
Trigger                           & Single-muon $\pt > 24/27/24\GeV$ \\
                                  & Single-electron $\pt > 27/32/32\GeV$ \\
Sign and flavor of leptons        & $\Pep\Pem$, $\PGmp\PGmm$, $\PGmpm\Pemp$ \\
Loose lepton count                & $=$2  \\
Tight lepton count                & $=$2 \\
Lepton $\pt$                      & $\pt > 26/29/26\GeV$  (muons)\\
                                  & $\pt > 29/34/34\GeV$  (electrons)\\
Tight lepton separation           & $\Delta R > 0.5$ \\
\PZ veto                          & $m_{\Pell\Pell} < 86$ or $> 96\GeV$ \\
AK8 jet threshold                 & $\pt>250 \GeV$ \\
AK8 jet count                     & $\geq$1 \\
\Htobb candidate                  & max $\Xtobb$ score AK8 jet \\
                                  & $70 < \MPN < 160\GeV$ \\
AK4 jet count for VBS             & $\geq$2 \\
VBS jet candidates                & max $\absdetajj$ \\
AK4 jet for \PQb tagging criteria & $\pt > 20\GeV$, $\abs{\eta}<2.5$ \\
Medium \PQb tagged jet count      & $=$0 \\
Missing transverse momentum       & $\ptmiss > 100\GeV$ \\
Event kinematics                  & $\ST > 500\GeV$ \\
\end{scotch}}
\end{table}

\begin{table}[!ht]
\centering
\topcaption{Selection criteria used in the OS \PZ dilepton channel. If the selection criteria differ across years, they are quoted as ``2016/2017/2018''.}
\label{tab:osz_presel}
\cmsTable{
\begin{scotch}{lc}
Selection                         & Description \\
\hline
Trigger                           & Single-muon $\pt > 24/27/24 \GeV$ \\
                                  & Single-electron $\pt > 27/32/32 \GeV$ \\
                                  & Dimuon $\pt > 17\,(8)\GeV$ \\
                                  & Dielectron $\pt > 23\,(12)\GeV$ \\
Sign and flavor of leptons        & $\Pep\Pem$, $\PGmp\PGmm$ \\
Loose lepton count                & $=$2  \\
Tight lepton count                & $=$2 \\
Leading lepton $\pt$              & $\pt > 19\,(25)\GeV$ muon (electron)\\
Subleading lepton $\pt$           & $\pt > 12\,(18)\GeV$ muon (electron) \\
\PZ window                        & $86 < m_{\Pell\Pell} < 96\GeV$  \\
AK8 jet threshold                 & $\pt>250/200\GeV$ \\
AK8 jet count                     & $\geq$2 \\
\Htobb candidate                  & max $\Xtobb$ score AK8 jet \\
                                  & $\Xtobb$ score  $>$0.2 \\
Boosted \Vtoqq candidate          & Other AK8 jet \\
                                  & $\Xtoqq$ score $>$0.6 \\
AK4 jet count for VBS             & $\geq$2 \\
VBS jet candidates                & max $\absdetajj$ \\
                                  & $\Mjj > 200\GeV$, $\absdetajj > 2$  \\
AK4 jet for \PQb tagging criteria & $\pt > 20\GeV$, $\abs{\eta}<2.5$ \\
Medium \PQb tagged jet count      & $=$0 \\
Event kinematics                  & $\ST > 500\GeV$ \\
                                  & $M_{\VVH} > 500\GeV$ \\
\end{scotch}}
\end{table}

The SS dilepton channel also uses mixed muon-electron triggers, requiring a muon (electron) with $\pt > 23\,(12)$\GeV and an electron (muon) with $\pt > 23\,(8)$\GeV, besides the single and dilepton triggers used in the OS dilepton channel.
Exactly two tight leptons are required, with no additional loose leptons. Both leptons are required to have $\pt > 30\GeV$, allowing at most one hadronically decaying $\PGt$ lepton.
At least one AK8 jet with $\pt > 250\GeV$ and two VBS jet candidates failing the medium \DeepJet working point are required.
If more than two VBS jet candidates are found, the same selection as in the single-lepton channel is applied.
To suppress events that are not consistent with VBS production, $\Mjj > 100\GeV$ is required.
The signal region is defined by $\Mjj > 500\GeV$, $\absdetajj > 3$, and \Xtobb score $>0.9$.
A control region includes events with $\Mjj > 100\GeV$ that fail the $\Mjj$ or $\detajj$ selections with respect to the signal region, with no requirement on the \Xtobb score of the selected AK8 jet.
The selection criteria for the SS dilepton channel are summarized in Table~\ref{tab:ss_presel}.

\begin{table}[!ht]
\centering
\topcaption{Selection criteria used in the SS dilepton channel. If the selection criteria differ across years, they are quoted as ``2016/2017/2018''.}
\label{tab:ss_presel}
\cmsTable{
\begin{scotch}{lc}
Selection                    & Description \\
\hline
Trigger                      & Single-muon $\pt > 24/27/24\GeV$ \\
                             & Single-electron $\pt > 27/32/32\GeV$ \\
                             & Dimuon $\pt > 17\,(8)\GeV$ \\
                             & Dielectron $\pt > 23\,(12)\GeV$ \\
                             & Electron (muon) $\pt > 23\,(8)\GeV$ \\
                             & Muon (electron) $\pt > 23\,(12)\GeV$ \\
Sign and flavor of leptons   &  $\Pepm\Pepm$, $\PGmpm\PGmpm$, $\Pepm\PGmpm$, $\PGmpm\PGtpm$, $\Pepm\PGtpm$ \\
Loose lepton count           & $=2$  \\
Tight lepton count           & $=2$ \\
Lepton $\pt$                 & $\pt>30\GeV$\\
AK8 jet threshold            & $\pt>250\GeV$ \\
AK8 jet count                & $\geq$1 \\
\Htobb candidate             & max $\Xtobb$ score AK8 jet \\
Medium \PQb tagged jet count & $=0$ \\
AK4 jet count for VBS        & $\geq2$ \\
VBS jet candidates           & max momentum jets, opposite $\eta$ sign \\
                             & $\Mjj > 100\GeV$  \\
\hline
Signal region                & $\Mjj > 500\GeV$, $\absdetajj > 3$, \\
                             & \Xtobb score $> 0.9$ for \Htobb candidate \\
\hline
Control region               & $\Mjj > 100\GeV$, \\
                             & with $\Mjj < 500\GeV$ or $\absdetajj < 3$ \\
\end{scotch}}
\end{table}

\subsection{Signal region definition}
\label{endmatter:mvaanalysis}

After the event selection, a multivariate analysis is performed in all channels.
In the all-hadronic, single-lepton and OS dilepton channels, an ``ABCD'' method~\cite{CDF:1990kbl} is used: the signal region, or region A, is defined alongside orthogonal control regions B, C, and D, used to predict the background in region A from data.
The selections used in each channel to define the signal region, or region A, are summarized in Table~\ref{tab:SRs}.
The ``AC axis'' selection, relying on the properties of the \VVH system, is inverted to define the region C; the ``AB axis'' selection, using the properties of the VBS jets, is inverted to define the region B; both are inverted to define the region D.
Additional selections, depending on the channel, are applied after the training of the discriminators to further improve the sensitivity of the search, and are common to the regions A, B, C and D.
These selections are applied only after the training in order to preserve the statistical power of the simulated samples at training time.
They are limited to the \ParticleNet \Xtobb and \Xtoqq scores, which are largely independent of the global event kinematic properties.

In the SS dilepton channel, a BDT is also trained to discriminate signal from background, and the signal region is divided into three bins according to the BDT output score ($\left[0, 0.1\right]$, $\left[0.1, 0.25\right]$, $\left[0.25, 1\right]$).

\subsection{Event yields}
\label{endmatter:yields}

Table~\ref{tab:yields} shows the data yields in the regions A, B, C and D of the all-hadronic, single-lepton, and OS dilepton channels, alongside the predicted background with its uncertainties, and the signal expected in region A for the $\kVV=2$ and SM benchmark scenarios.
The background is predicted to be small in region A for all channels, except for the all-hadronic semi-boosted channel, where the contribution of the QCD multijet background is larger.
Systematic uncertainties in the background predictions are derived from closure tests in data.
The systematic uncertainties in the signal yields correspond to the sum in quadrature of all the experimental and theoretical uncertainties.
The statistical unceratainties arise from the limited size of the simulated samples for signal, and from the amount of data in regions B, C and D for the background.

Table~\ref{tab:yields_ss} shows the data yields and the predicted background in the signal (SR) and control regions (CR) of the same-sign dilepton channel for the $\Pell\Pell$, where $\Pell=\Pe,\PGm$, and the $\Pell\PGt$ categories, alongside with the expected yields for the benchmark SM signal.
The total uncertainty in the background is shown, which accounts for both statistical and systematic sources.

\ifthenelse{\boolean{cms@external}}{\newpage}{}
\begin{table*}[htpb]
\centering
\topcaption{
Summary of the selections used to define the signal region, or region A, in the all-hadronic, single-lepton, and OS dilepton channels.
The selections on the ``AB axis'' and ``AC axis'' are inverted to define the regions B, C and D.
Additional selections on the \Xtobb and \Xtoqq scores are applied after the training to further improve the sensitivity of the search.
For the all-hadronic fully boosted channel, the selections on the \Xtoqq score are applied on the \pt-leading/subleading vector boson candidates.
}
\label{tab:SRs}
\cmsTableu{
\begin{scotch}{cccccc}
               & \multicolumn{2}{c}{All-hadronic}     & Single-lepton & \multicolumn{2}{c}{Dilepton} \\
               & fully boosted    & semi-boosted      &               & OS \PW\PW          & OS \PZ \\
\hline
AC axis (\VVH) & DNN $>0.97$      & DNN $>0.8$        & DNN $>0.9$    & BDT $>0.85$        & BDT $>0.85$ \\
\hline
AB axis (VBS)  & $\absdetajj > 5$ & $\absdetajj > 6$  & BDT $>0.82$   & $\absdetajj > 4.9$ & $\absdetajj > 4$ \\
               &                  & $\Mjj > 1300\GeV$ &               & $\Mjj > 760\GeV$   & $\Mjj > 750\GeV$ \\
\hline
\Xtobb score   & $>$0.8           & \NA               & $>$0.5        & $>$0.7             & \NA \\
\Xtoqq score   & $>$0.8/0.7       & \NA               & $>$0.7        & \NA                & \NA  \\
\end{scotch}
}
\end{table*}

\begin{table*}[htpb]
\centering
\topcaption{
Data yields in regions B, C and D, used to estimate the background in region A, along with data, predicted background, and expected signal in region A.
Signal yields are shown for the SM and $\kVV=2$ benchmarks.
}
\label{tab:yields}
\cmsTableu{\begin{scotch}{lcccccc}
              & \multicolumn{2}{c}{All-hadronic}                             & Single-lepton                & \multicolumn{2}{c}{Dilepton} \\
              & fully boosted                & semi-boosted                  &                              & OS \PW\PW                    & OS \PZ \\
\hline
B data        & 1                            & 410                           & 3                            & 2                            & 1 \\
C data        & 73                           & 399                           & 61                           & 10                           & 5 \\
D data        & 654                          & 15\,974                       & 1039                         & 179                          & 17 \\
\hline
A data        & 0                            & 7                             & 0                            & 0                            & 0 \\
A prediction  & $0.11^{+0.26}_{-0.09}\stat $ & $10.24^{+0.78}_{-0.72}\stat $ & $0.18^{+0.18}_{-0.10}\stat $ & $0.11^{+0.18}_{-0.08}\stat $ & $0.29^{+0.91}_{-0.25}\stat $ \\
              & $\pm 0.03\syst$              & $\pm 1.95\syst$               & $\pm 0.06\syst$              & $\pm 0.02\syst$              & $\pm 0.21\syst$ \\
A signal      & $3.30 \pm 0.02\stat $        & $1.06 \pm 0.01\stat $         & $2.79 \pm 0.02\stat $        & $0.31 \pm 0.01\stat $        & $0.22 \pm (<0.01)\stat $ \\
($\kVV=2$)    & $\pm 1.01\syst$              & $\pm 0.36\syst$               & $\pm 0.84\syst$              & $\pm 0.10\syst$              & $\pm 0.07\syst$ \\
A signal (SM) & $0.04 \pm 0.01$              & $0.01 \pm ({<}0.01)$            & $0.04 \pm 0.02$              & $0.01 \pm ({<}0.01)$           & $({<}0.01) \pm ({<}0.01)$ \\
\end{scotch}}
\end{table*}

\begin{table*}[htpb]
\centering
\topcaption{
Data yields, predicted background and expected signal in signal (SR) and control regions (CR) of the same-sign dilepton channel.
}
\label{tab:yields_ss}
\cmsTableu{
\begin{scotch}{lcccc}
            & \multicolumn{4}{c}{Same-sign dilepton} \\
            & $\Pell\Pell$ SR      & $\Pell\Pell$ CR         & $\Pell\PGt$ SR         & $\Pell\PGt$ CR \\
\hline
Data        & 4                  & 39                    & 3                     & 49\\
Background  & $1.90 \pm 0.33$    & $27.90 \pm 3.67$      & $3.46 \pm 0.55$       & $50.52 \pm 8.09$ \\
Signal (SM) & $0.01 \pm (<0.01)$ & $({<}0.01) \pm (<0.01)$ & $({<}0.01) \pm (<0.01)$ & $({<}0.01) \pm ({<}0.01)$ \\
\end{scotch}
}
\end{table*}

\ifthenelse{\boolean{cms@external}}
{}
{
\clearpage
\numberwithin{figure}{section}
\numberwithin{table}{section}
\section{Supplemental Material}
\label{suppmat}
\subsection{Neural network event classifiers}
\label{supp:dnn}

The results presented in this paper exploit deep neural network (DNN) classifiers to
select events and define signal and control regions.
Below we present additional studies on these DNNs and on the definition of the ``ABCD'' samples that are used to estimate the backgrounds in the all-hadronic, single-lepton and OS dilepton channels of the analysis.

Figures~\ref{fig:2dallhaddata}~and~\ref{fig:2dallhad} (Figs.~\ref{fig:2dslepdata}~and~\ref{fig:2dslep}) demonstrate the statistical independence of the variables used to define the ABCD axes in the all-hadronic (single-lepton) channel, in data and in simulated events, respectively.

\begin{figure}[tb]
\centering
\includegraphics[width=\cmsFigWidth]{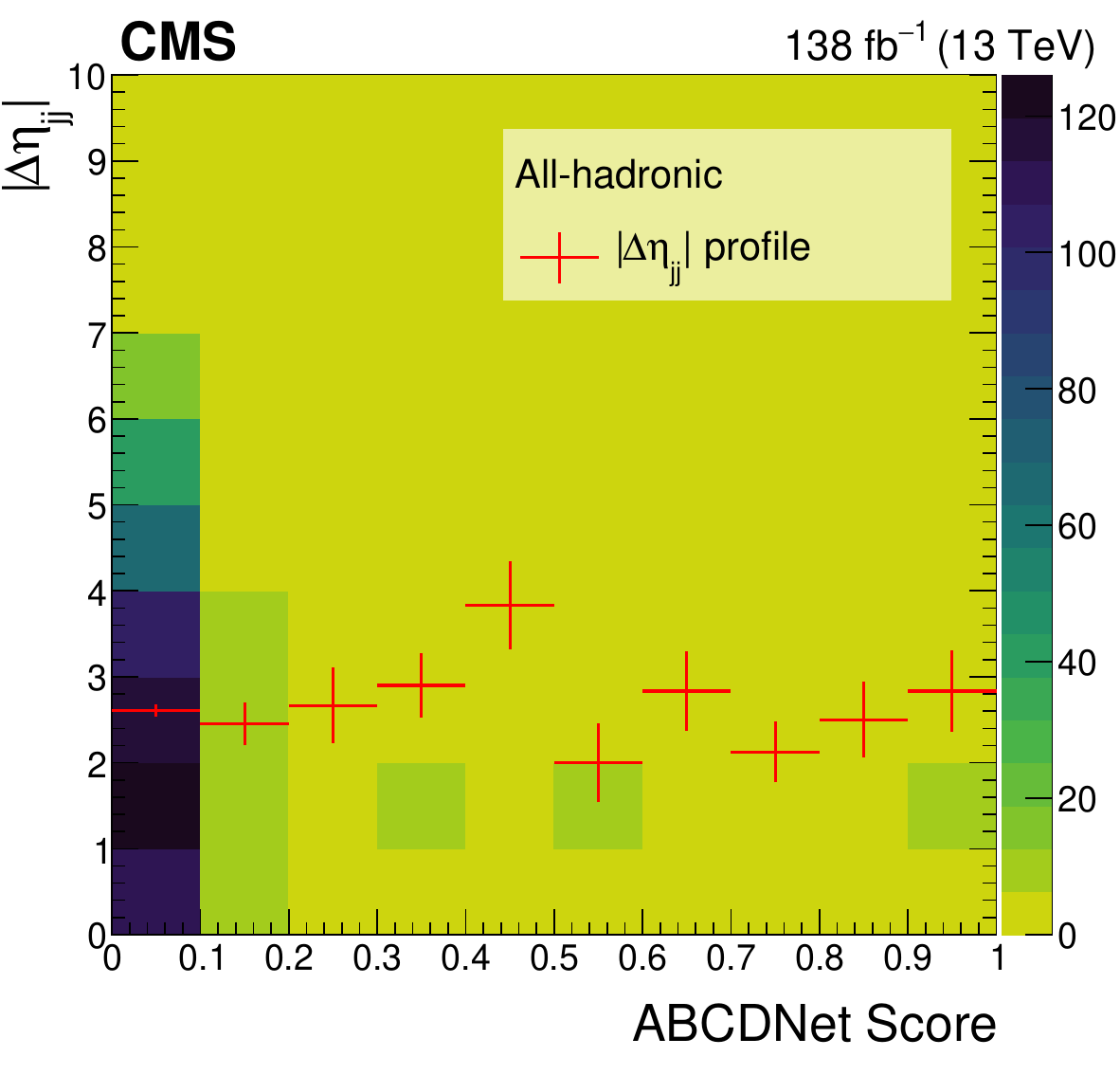}
\caption{
Two-dimensional distribution of the DNN output and the \absdetajj in the all-hadronic channel, as obtained in data. These are the axes used in this channel to define the ABCD samples.
A profile histogram is overlaid to better depict the statistical independence of the two variables.}
\label{fig:2dallhaddata}
\end{figure}

\begin{figure}[htbp!]
\centering
\includegraphics[width=\cmsFigWidth]{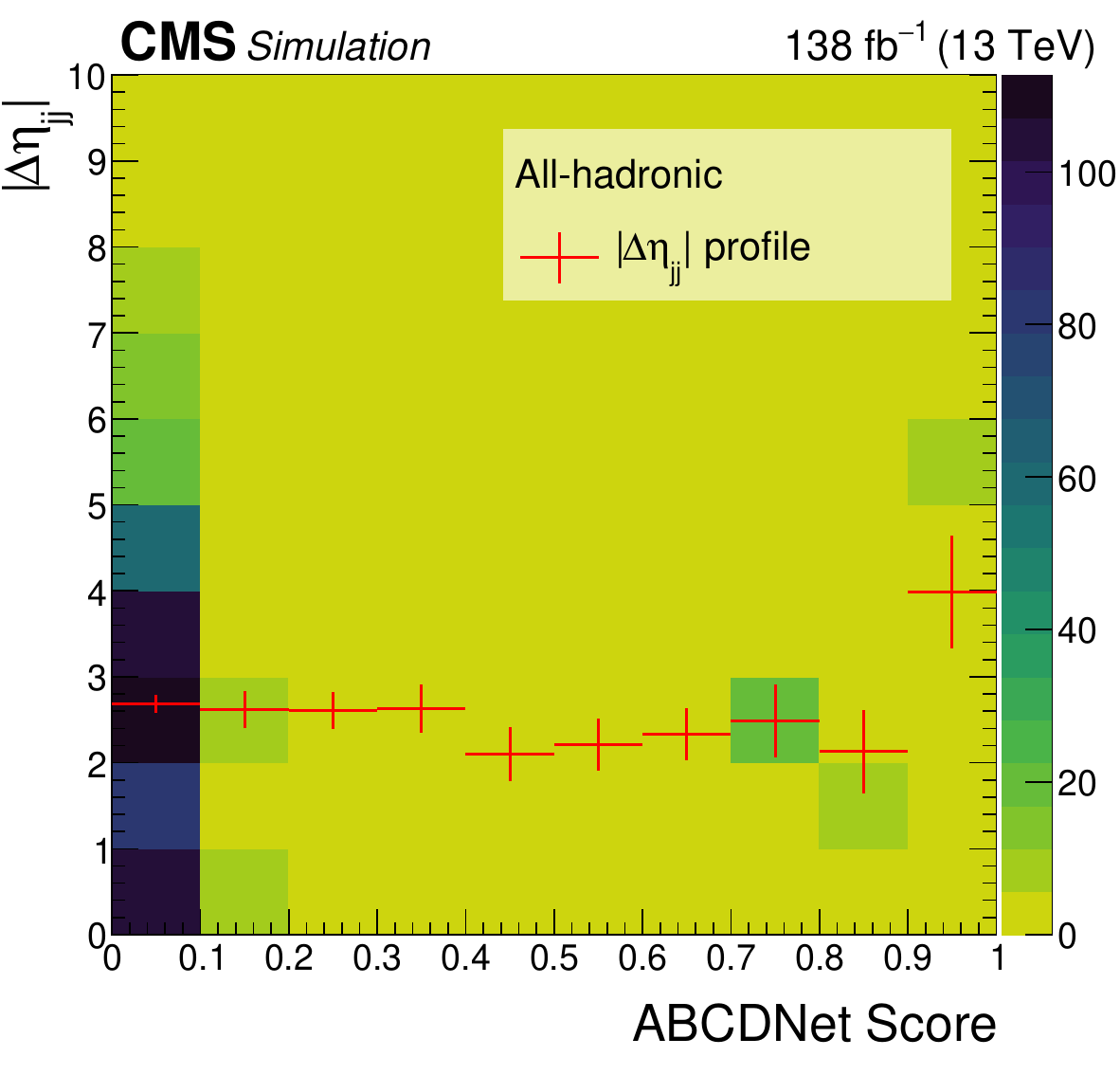}
\caption{
Two-dimensional distribution of the DNN output and the \absdetajj in the all-hadronic channel, as obtained in background simulated events. These are the axes used in this channel to define the ABCD samples.
A profile histogram is overlaid to better depict the statistical independence of the two variables.}
\label{fig:2dallhad}
\end{figure}

\begin{figure}[htbp!]
\centering
\includegraphics[width=\cmsFigWidth]{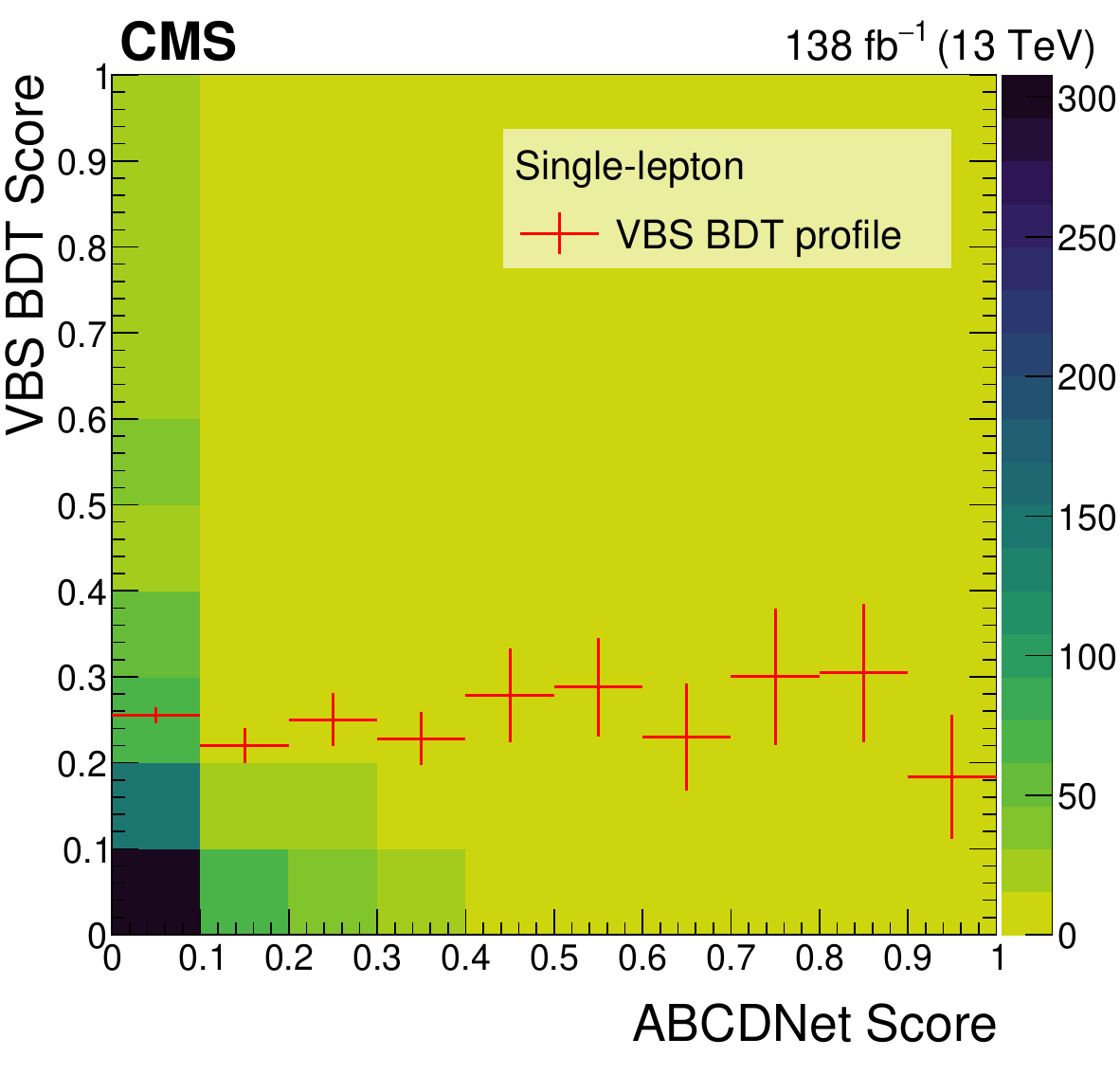}
\caption{
Two-dimensional distribution of the DNN output and the VBS BDT output in the single-lepton channel, as obtained in data. These are the axes used in this channel to define the ABCD samples.
A profile histogram is overlaid to better depict the statistical independence of the two variables.}
\label{fig:2dslepdata}
\end{figure}

\begin{figure}[htbp!]
\centering
\includegraphics[width=\cmsFigWidth]{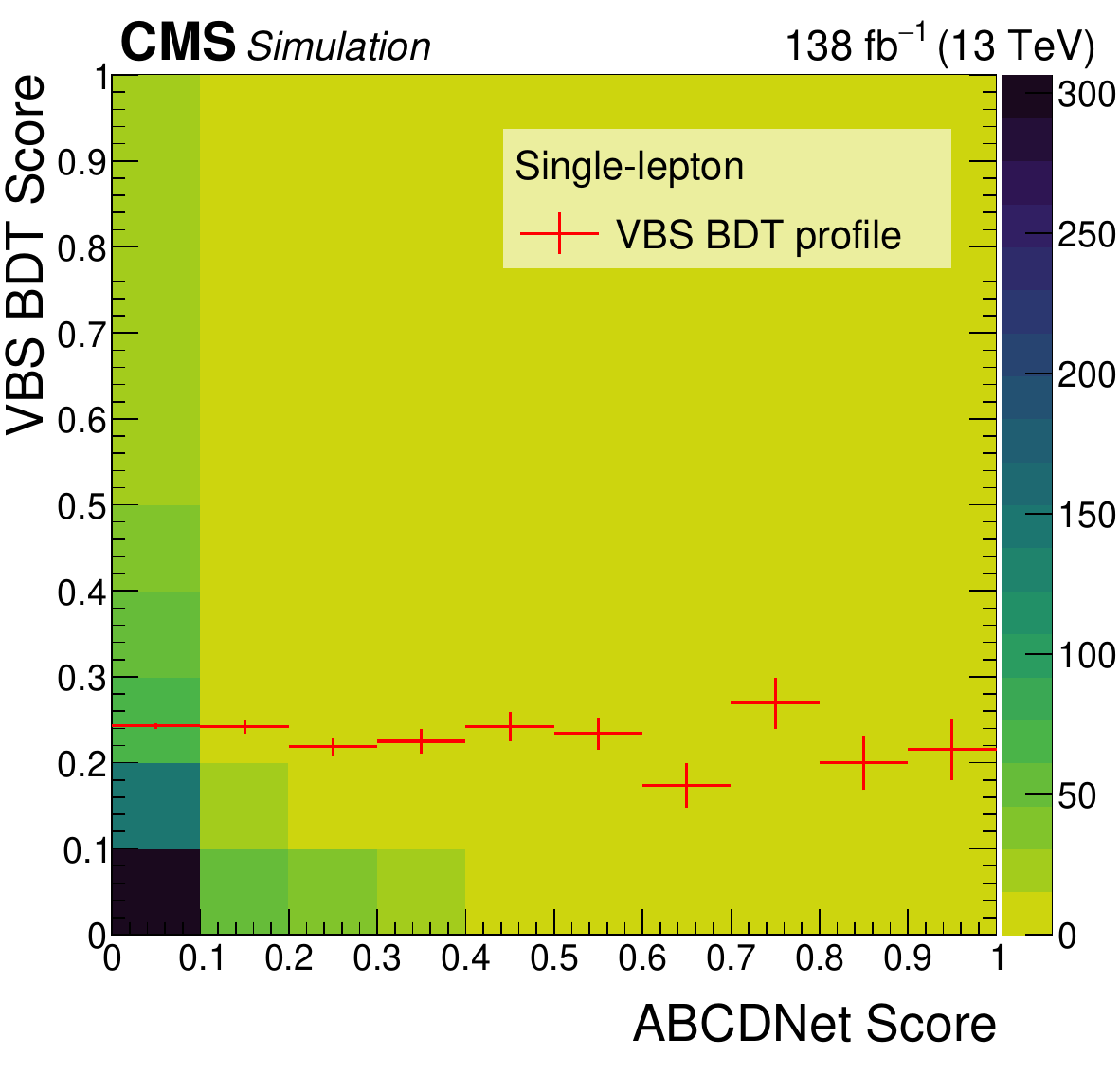}
\caption{
Two-dimensional distribution of the DNN output and the VBS BDT output in the single-lepton channel, as obtained in background simulated events. These are the axes used in this channel to define the ABCD samples.
A profile histogram is overlaid to better depict the statistical independence of the two variables.}
\label{fig:2dslep}
\end{figure}

Figures~\ref{fig:allhad_dnn1}~and~\ref{fig:allhad_dnn2} (figures~\ref{fig:slep_dnn1}~and~\ref{fig:slep_dnn2}) show the distributions of the DNN output in the regions B and D and in the regions A and C of the all-hadronic (single-lepton) channel, respectively.

\begin{figure}[htbp!]
   \centering
   \includegraphics[width=\cmsFigWidth]{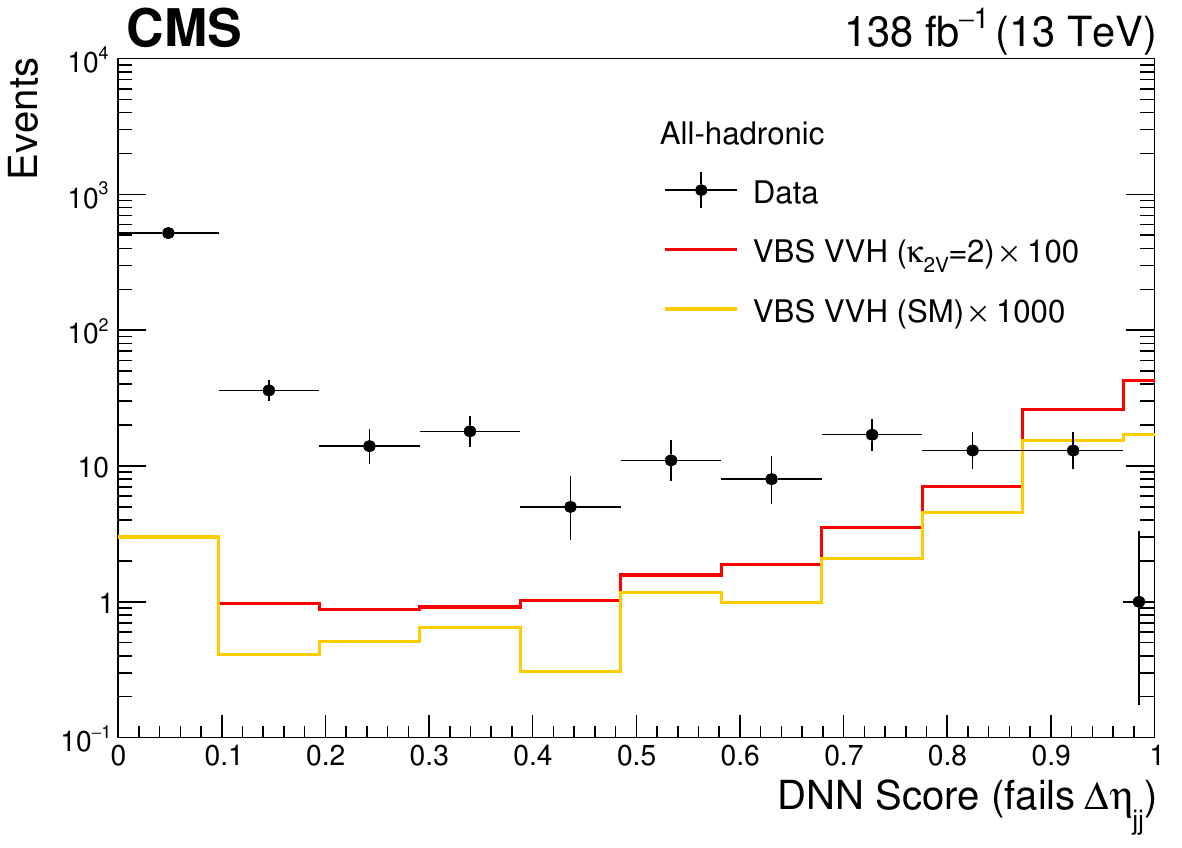}
	\caption{
        Distribution of the DNN output in the regions B and D of the all-hadronic channel.
        The expected signals for the standard model scenario (orange) and for $\kVV = 2$ (red)
        are superimposed to the data (black markers).
        The cross section used for signal is computed at leading order using \MADGRAPH,
        and is scaled by a factor 1000 (100) for the SM ($\kVV = 2$) signal to improve visibility.}
   \label{fig:allhad_dnn1}
\end{figure}

\begin{figure}[htbp!]
\centering
\includegraphics[width=\cmsFigWidth]{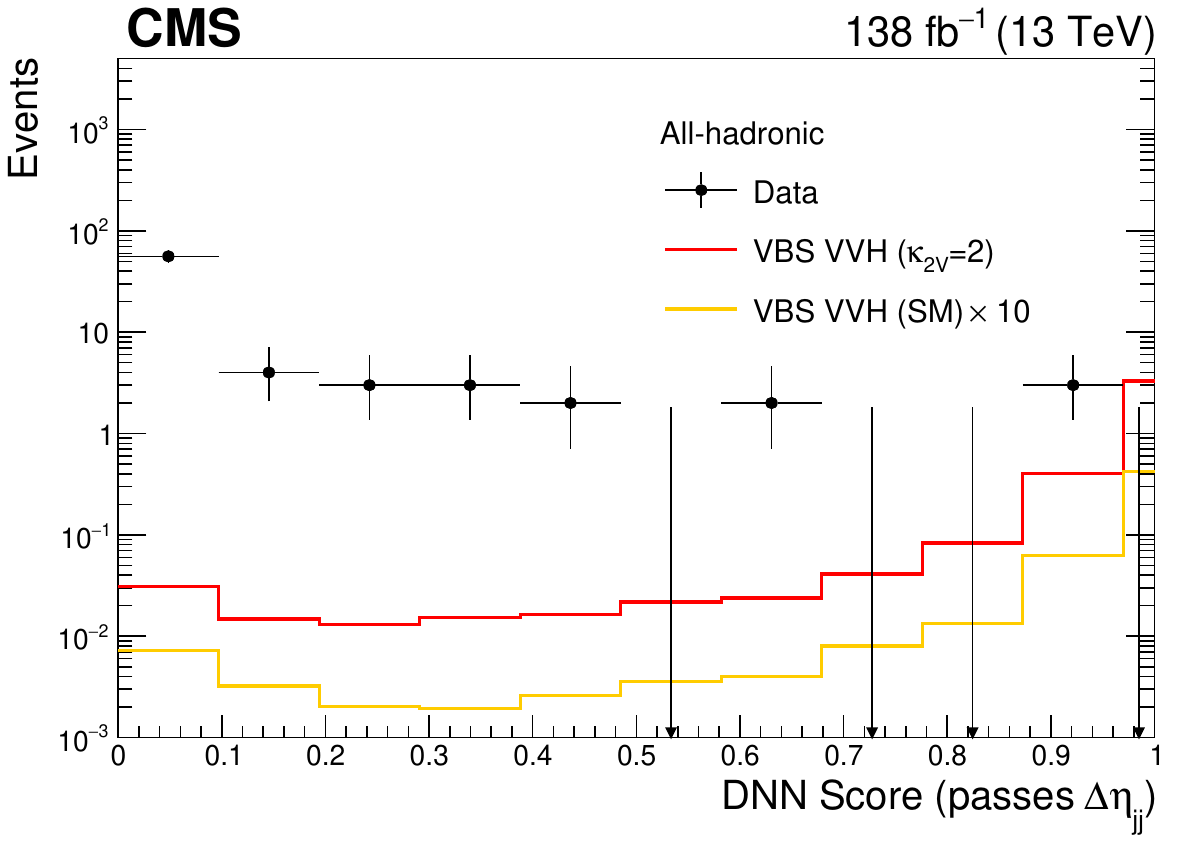}
\caption{
Distribution of the DNN output in the regions A and C of the all-hadronic channel.
The expected signals for the standard model scenario (orange) and for $\kVV = 2$ (red) are superimposed to the data (black markers).
The cross section used for signal is computed at leading order using \MADGRAPH, and is scaled by a factor 10 for the SM signal to improve visibility.}
\label{fig:allhad_dnn2}
\end{figure}

\begin{figure}[htbp!]
\centering
\includegraphics[width=\cmsFigWidth]{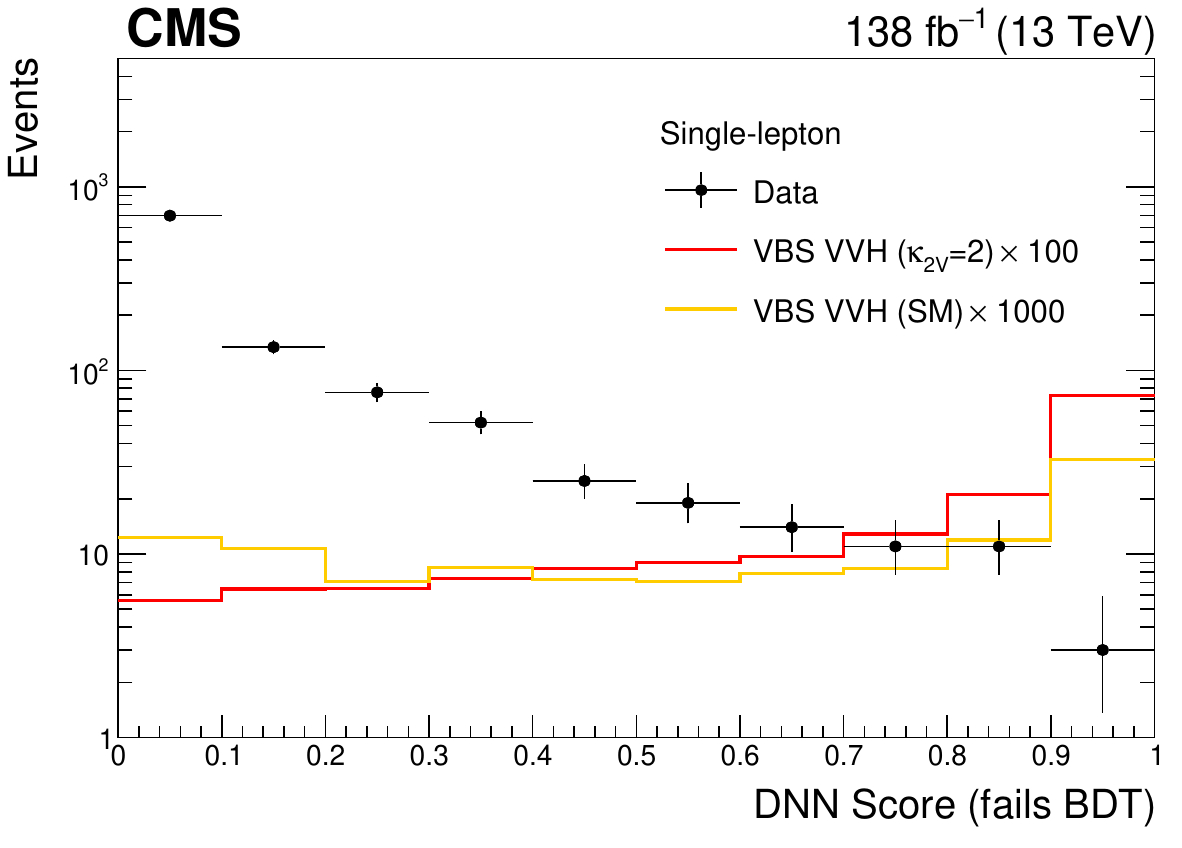}
\caption{
Distribution of the DNN output in the regions B and D of the single-lepton channel.
The expected signals for the standard model scenario (orange) and for $\kVV = 2$ (red) are superimposed to the data (black markers).
The cross section used for signal is computed at leading order using \MADGRAPH, and is scaled by a factor 1000 (100) for the SM ($\kVV = 2$) signal to improve visibility.}
\label{fig:slep_dnn1}
\end{figure}

\begin{figure}[htbp!]
\centering
\includegraphics[width=\cmsFigWidth]{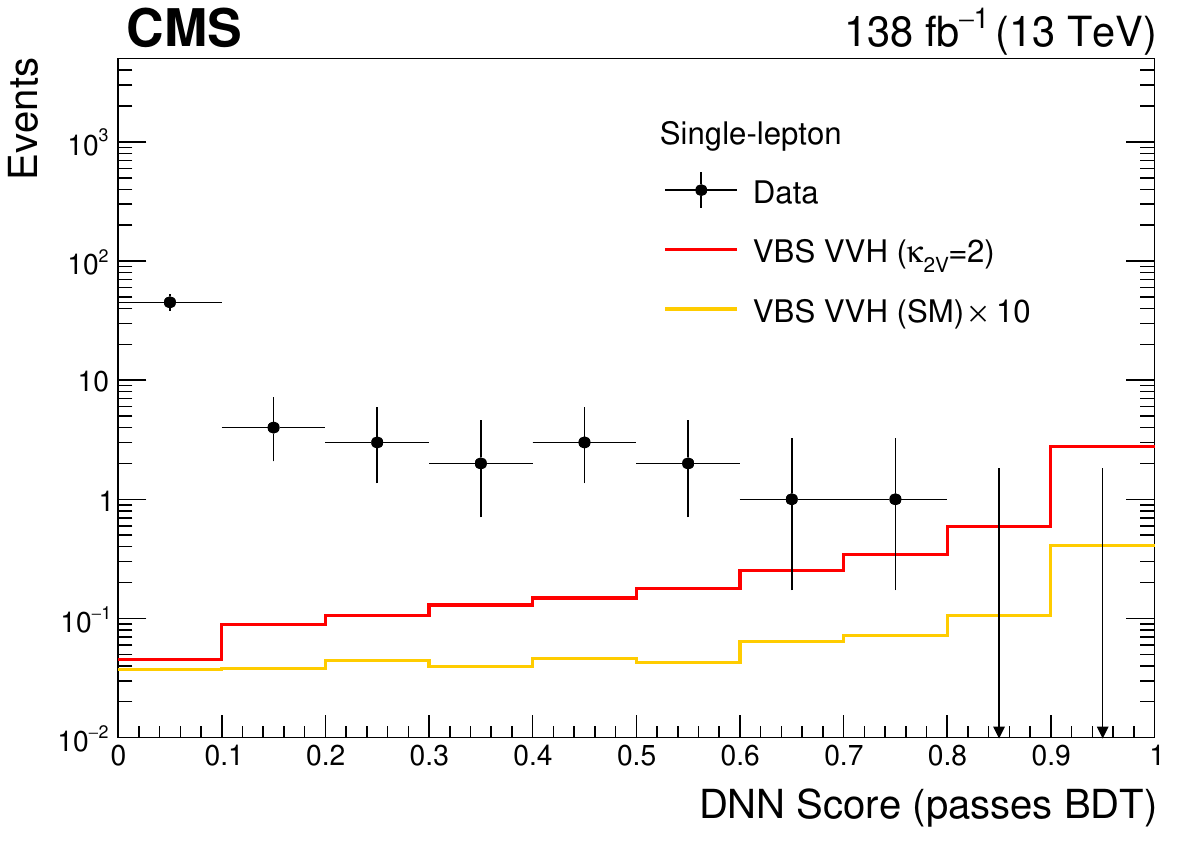}
\caption{
Distribution of the DNN output in the regions A and C of the single-lepton channel.
The expected signals for the standard model scenario (orange) and for $\kVV = 2$ (red) are superimposed to the data (black markers).
The cross section used for signal is computed at leading order using \MADGRAPH, and is scaled by a factor 10 for the SM signal to improve visibility.}
\label{fig:slep_dnn2}
\end{figure}
\ifthenelse{\boolean{cms@external}}{\clearpage}{}

\subsection{Analysis sensitivity: additional studies}
\label{supp:channels}

The results presented in the paper are a combination of the six analysis channels.
Below we present results obtained in individual channels, including comparisons to those arising from their combination.

Figures~\ref{fig:k2wexpobs}~and~\ref{fig:k2zexpobs} show the 95\%~\CL expected and observed upper limits on the VBS \VVH production cross section as a function of the \kVV coupling modifier for each analysis channel.
The combined 95\%~\CL upper limits are also shown.

\begin{figure}[htbp!]
\centering
\includegraphics[width=\cmsFigWidth]{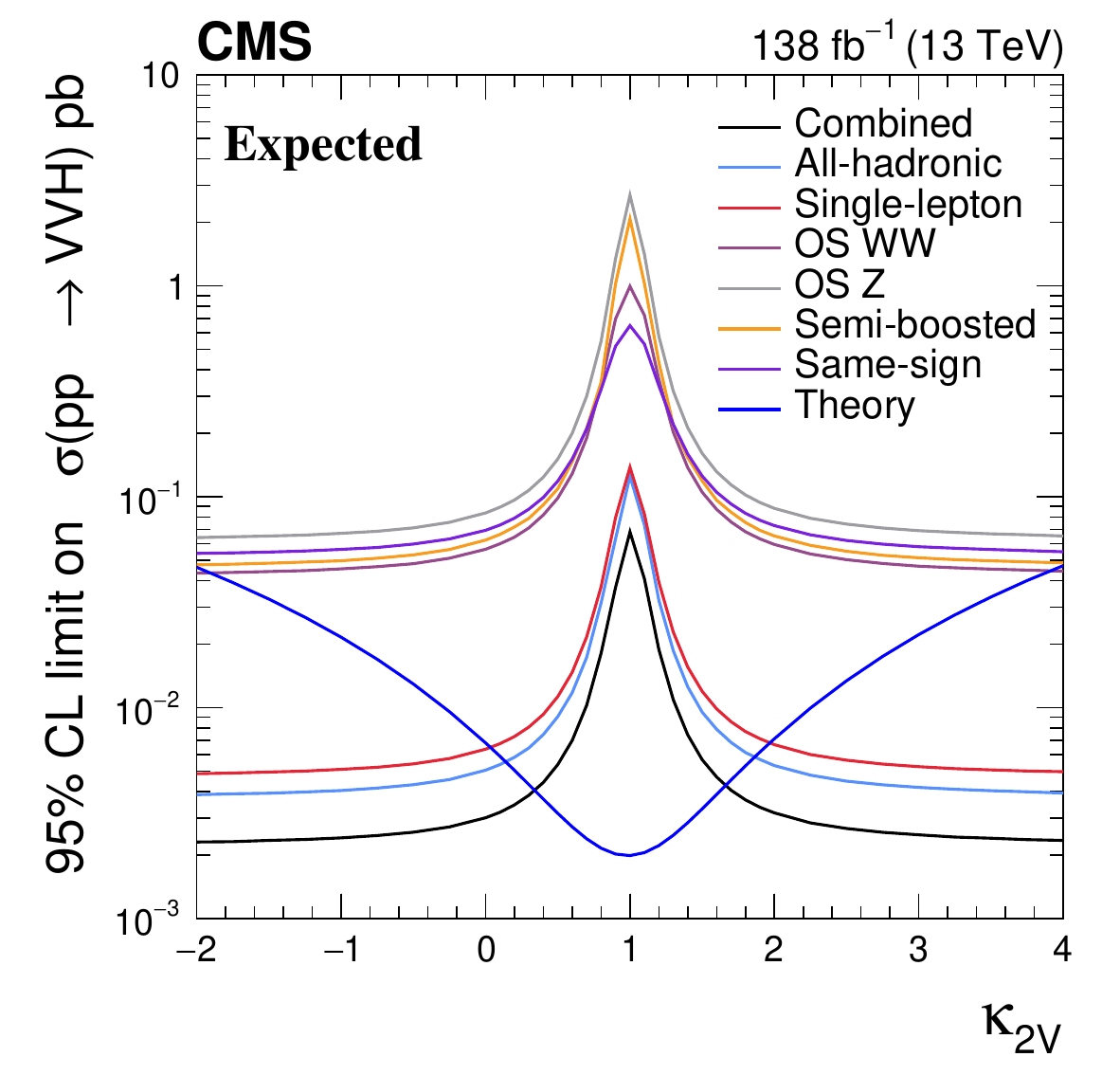}
\includegraphics[width=\cmsFigWidth]{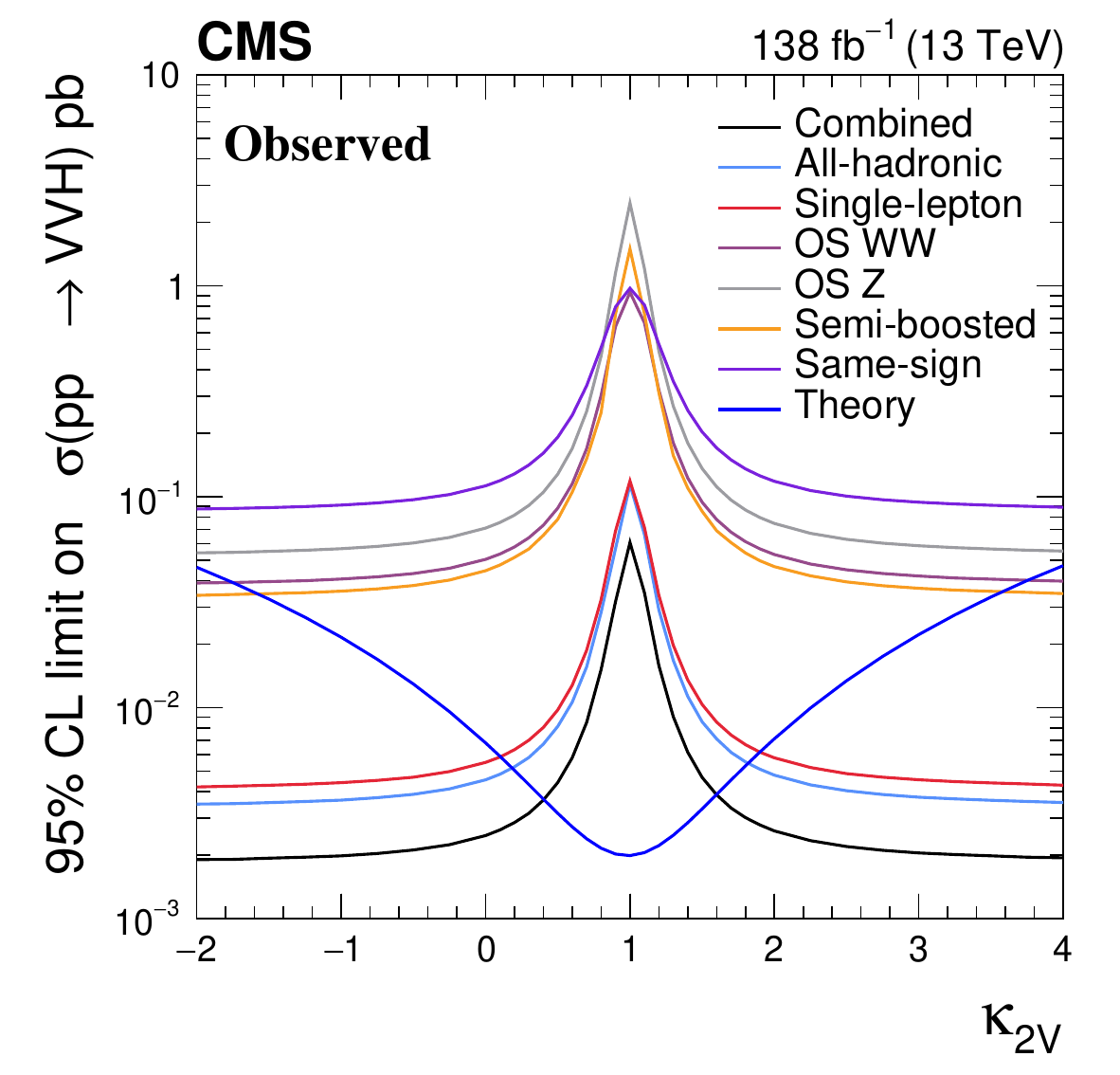}
\caption{Comparison of the expected (\cmsLeft) and observed (\cmsRight) 95\%~\CL limits on \kVV.
The combined limit is shown as in black, while the limits obtained in individual channels are shown as colored lines.}
\label{fig:k2vexpobs}
\end{figure}

Figures~\ref{fig:k2wexpobs}~and~\ref{fig:k2zexpobs} show the 95\%~\CL upper limits on the VBS \VVH production cross section as a function of the \kWW and \kZZ coupling modifiers
for each analysis channel.
The expected and observed upper limits are shown in Fig.~\ref{fig:k2wexpobs} as a function of \kWW. The upper limits as a function of \kZZ are shown in Fig.~\ref{fig:k2zexpobs}.
The combined 95\%~\CL upper limits are also shown.

\begin{figure}[htbp!]
\centering
\includegraphics[width=\cmsFigWidth]{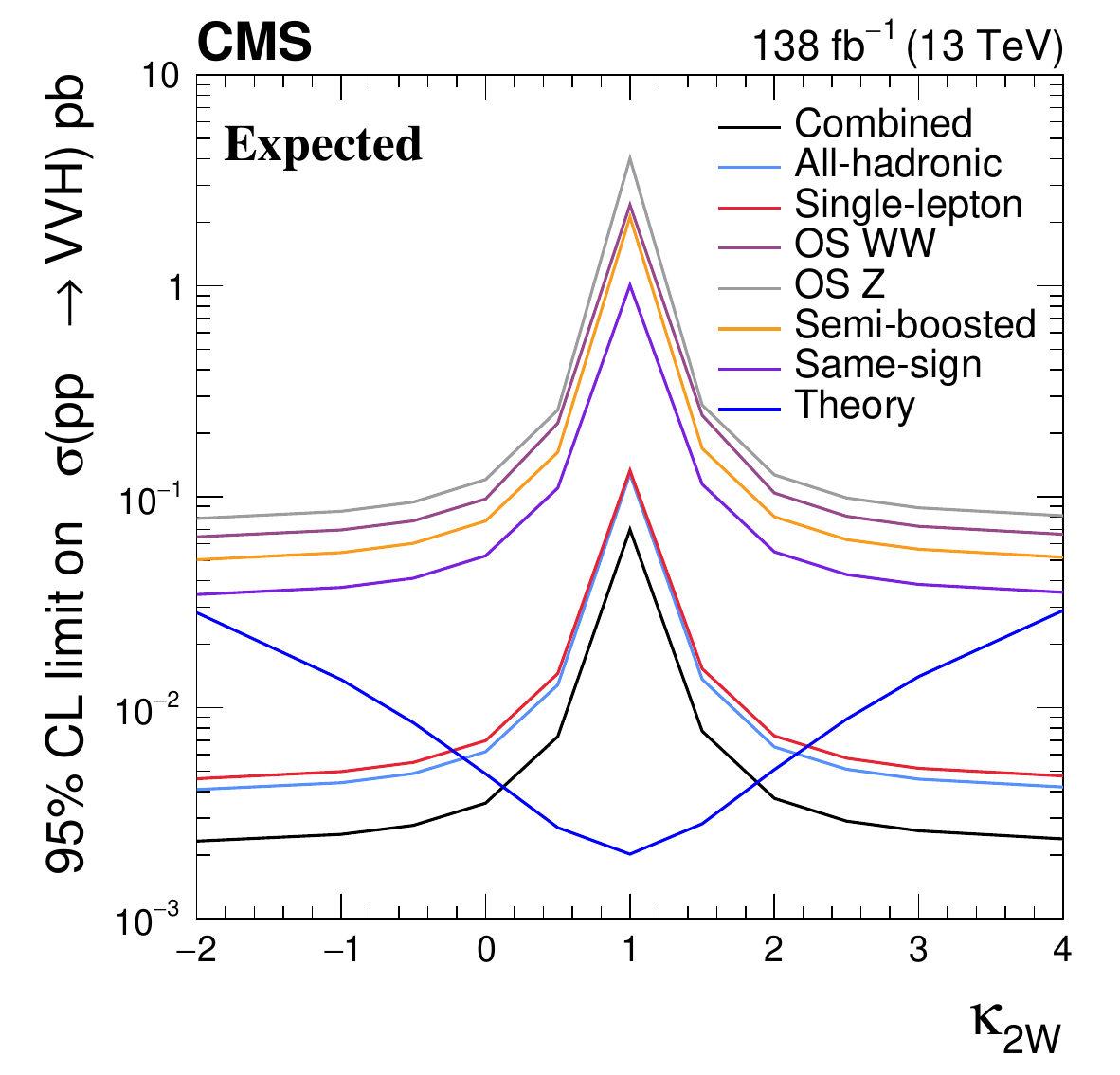}
\includegraphics[width=\cmsFigWidth]{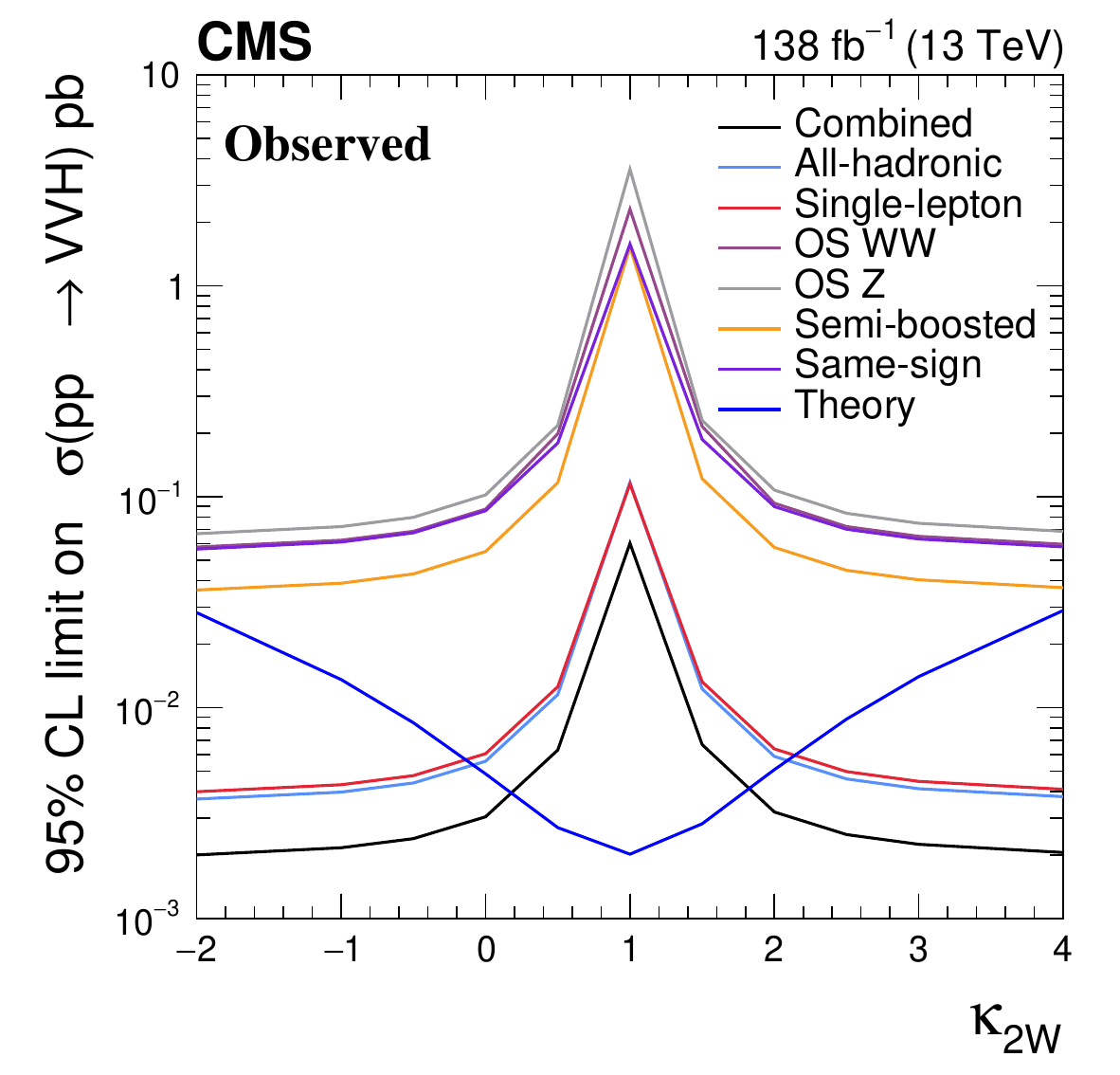}
\caption{Comparison of the expected (\cmsLeft) and observed (\cmsRight) 95\%~\CL limits on \kWW.
The combined limit is shown as in black, while the limits obtained in individual channels are shown as colored lines.}
\label{fig:k2wexpobs}
\end{figure}

\begin{figure}[htbp!]
\centering
\includegraphics[width=\cmsFigWidth]{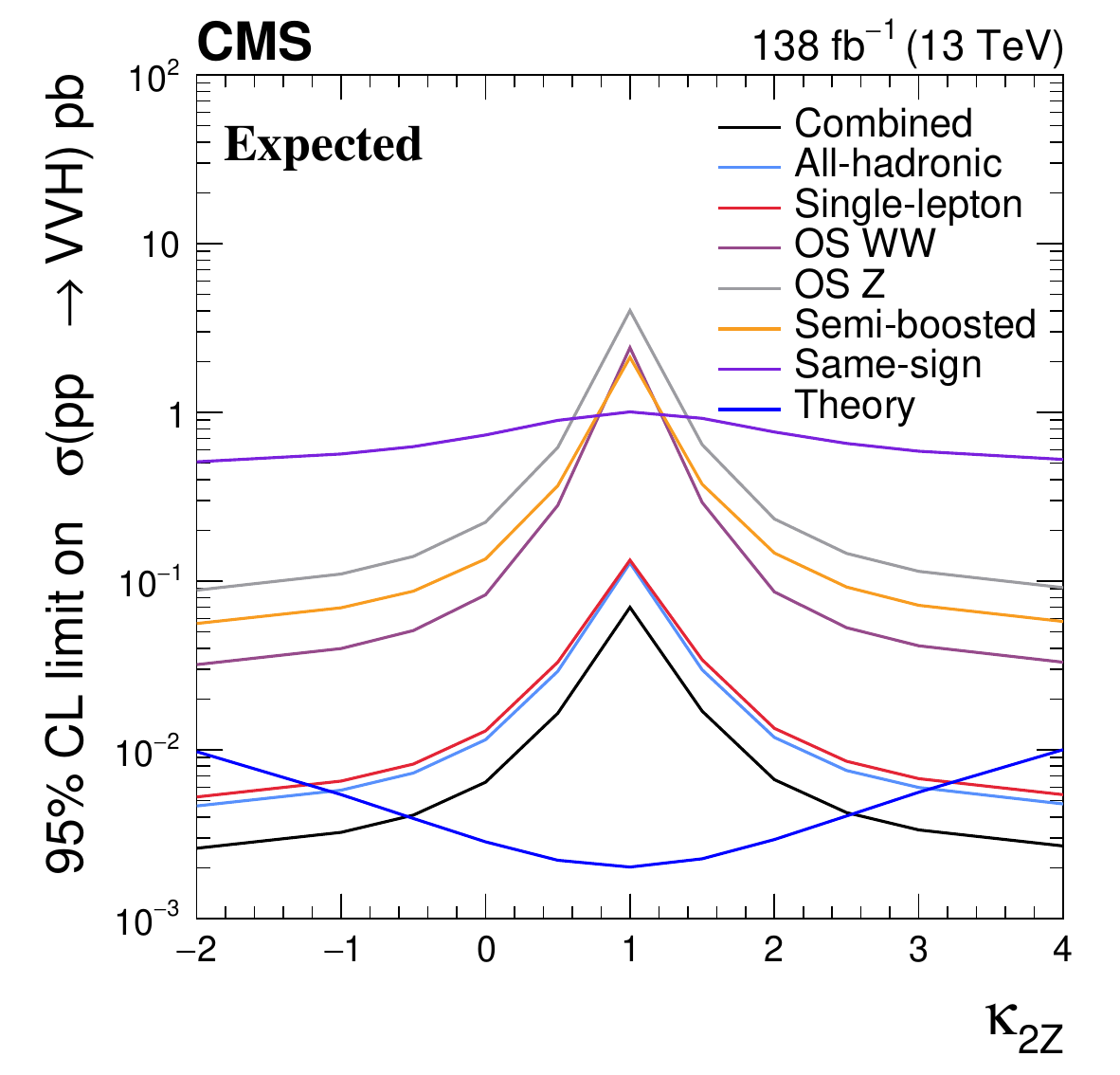}
\includegraphics[width=\cmsFigWidth]{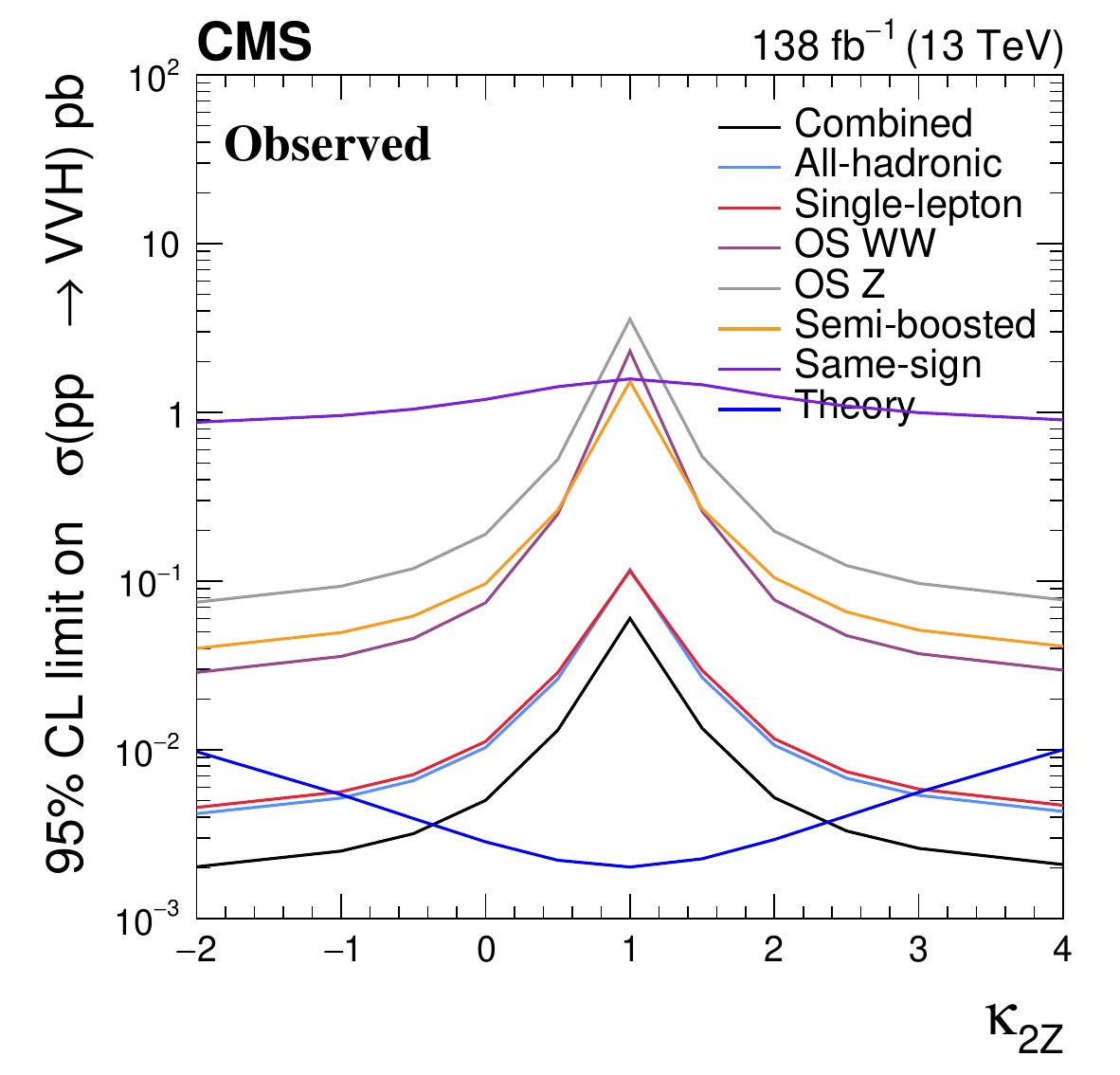}
\caption{Comparison of the expected (\cmsLeft) and observed (\cmsRight) 95\%~\CL limits on \kZZ.
The combined limit is shown as in black, while the limits obtained in individual channels are shown as colored lines.}
\label{fig:k2zexpobs}
\end{figure}

Figure~\ref{fig:2dexpobs} shows the two-dimensional expected and observed exclusion regions in the \kWW-\kZZ plane, corresponding to 2 standard deviations (\std), as obtained in each analysis channel.
The exclusion regions obtained from the combination of all channels are also shown.

\begin{figure}[htbp!]
\centering
\includegraphics[width=\cmsFigWidth]{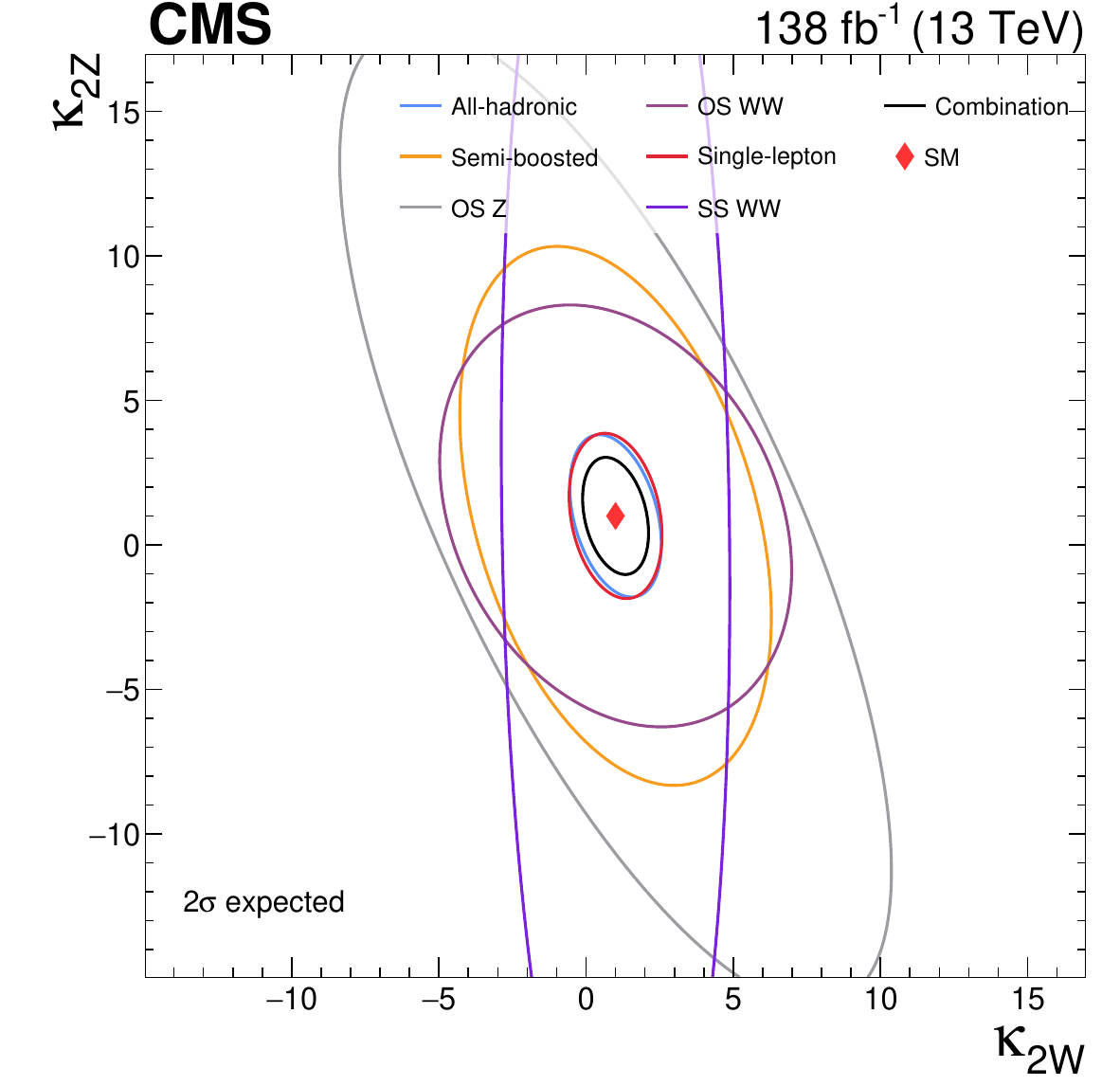}
\includegraphics[width=\cmsFigWidth]{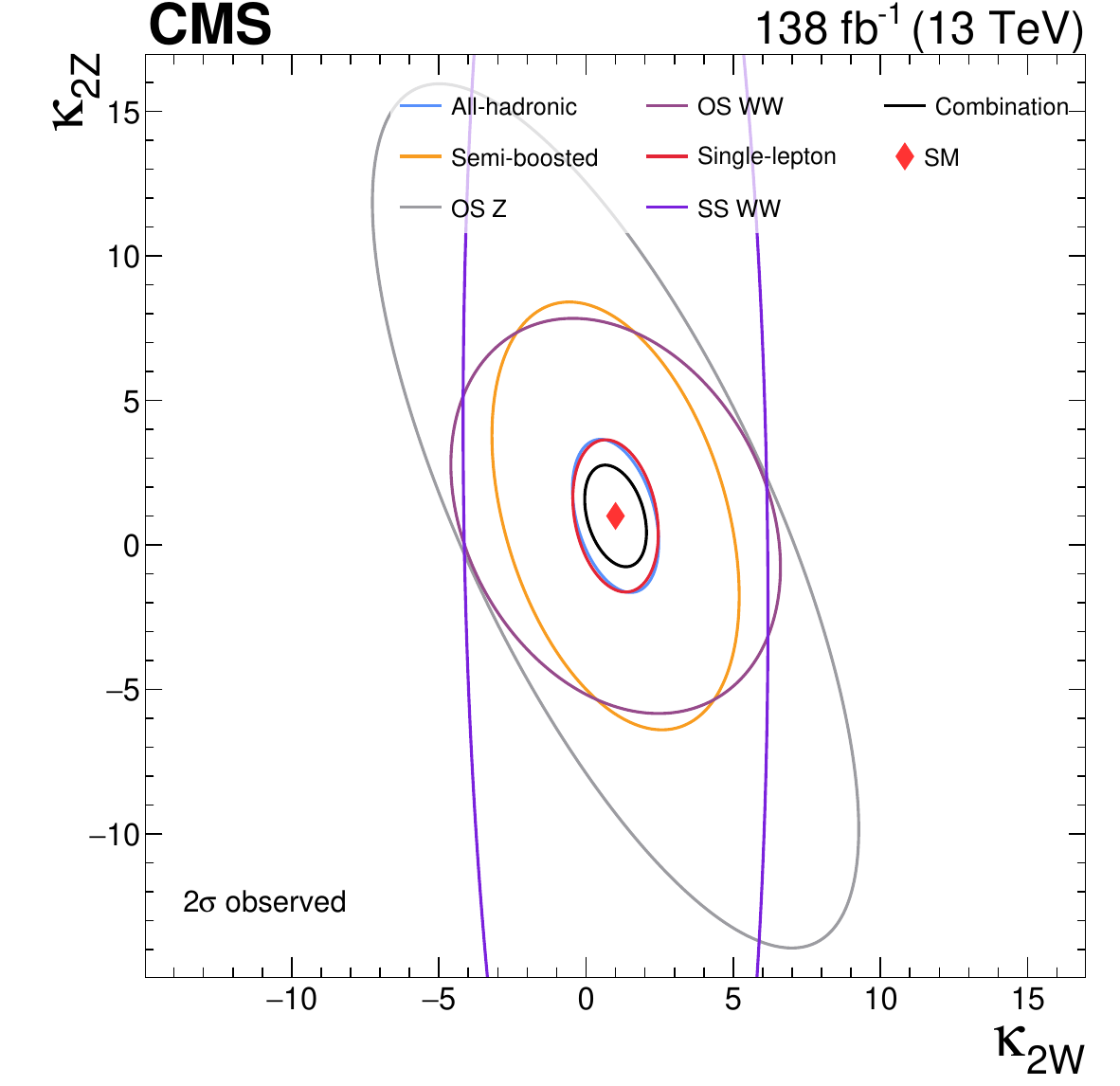}
\caption{Comparison of the expected (\cmsLeft) and observed (\cmsRight) exclusion regions
corresponding to 2 standard deviations (\std), in the two-dimensional \kWW-\kZZ plane.
The combined exclusion region is shown as in black, while the exclusion regions obtained in individual channels are shown as colored lines.}
\label{fig:2dexpobs}
\end{figure}

}
\cleardoublepage \section{The CMS Collaboration \label{app:collab}}\begin{sloppypar}\hyphenpenalty=5000\widowpenalty=500\clubpenalty=5000\cmsinstitute{Yerevan Physics Institute, Yerevan, Armenia}
{\tolerance=6000
A.~Hayrapetyan, V.~Makarenko\cmsorcid{0000-0002-8406-8605}, A.~Tumasyan\cmsAuthorMark{1}\cmsorcid{0009-0000-0684-6742}
\par}
\cmsinstitute{Institut f\"{u}r Hochenergiephysik, Vienna, Austria}
{\tolerance=6000
W.~Adam\cmsorcid{0000-0001-9099-4341}, L.~Benato\cmsorcid{0000-0001-5135-7489}, T.~Bergauer\cmsorcid{0000-0002-5786-0293}, M.~Dragicevic\cmsorcid{0000-0003-1967-6783}, C.~Giordano\cmsorcid{0000-0001-6317-2481}, P.S.~Hussain\cmsorcid{0000-0002-4825-5278}, M.~Jeitler\cmsAuthorMark{2}\cmsorcid{0000-0002-5141-9560}, N.~Krammer\cmsorcid{0000-0002-0548-0985}, A.~Li\cmsorcid{0000-0002-4547-116X}, D.~Liko\cmsorcid{0000-0002-3380-473X}, M.~Matthewman, J.~Schieck\cmsAuthorMark{2}\cmsorcid{0000-0002-1058-8093}, R.~Sch\"{o}fbeck\cmsAuthorMark{2}\cmsorcid{0000-0002-2332-8784}, M.~Shooshtari\cmsorcid{0009-0004-8882-4887}, M.~Sonawane\cmsorcid{0000-0003-0510-7010}, W.~Waltenberger\cmsorcid{0000-0002-6215-7228}, C.E.~Wulz\cmsAuthorMark{2}\cmsorcid{0000-0001-9226-5812}
% ...
\par}
$^{1}$Also at Yerevan State University, Yerevan, Armenia\\
$^{2}$Also at Technische Universit\"{a}t Wien, Vienna, Austria\\
% ...
\end{sloppypar}
\end{document}